\documentclass[aps,prab,reprint,superscriptaddress,%
amsmath,amssymb,longbibliography,floatfix]{revtex4-2}
\usepackage[T1]{fontenc}
\usepackage{graphicx,bm,booktabs,microtype,hyperref}
\usepackage{xcolor}
\hypersetup{hidelinks,pdftitle={Coherent synchrotron radiation extraction from a
  bent beam pipe}}
\graphicspath{{./}}
\newcommand{\dd}{\mathrm d}
\newcommand{\ii}{\mathrm i}
\newcommand{\Z}{Z_0}

\begin{document}
\title{Coherent synchrotron radiation extraction from a bent beam pipe}
\author{Igor Zagorodnov}
\email{igor.zagorodnov@desy.de}
\affiliation{Deutsches Elektronen-Synchrotron DESY, Notkestra{\ss}e 85,
22607 Hamburg, Germany}
\author{Dmitry Bazyl}
\affiliation{Deutsches Elektronen-Synchrotron DESY, Notkestra{\ss}e 85,
22607 Hamburg, Germany}
\author{Gennady Stupakov}
\affiliation{xLight Inc., Palo Alto, California 94306, USA}
\date{\today}
\begin{abstract}
Coherent synchrotron radiation generated in a bending
magnet can propagate far downstream and heat the beam pipe.
We investigate a beam-free side outlet that diverts part
of this radiation into a separate guide.
We use a paraxial equation to calculate radiation generation in the bend
and its division between the two guides. Modal and finite-difference
solutions are used in the bend, and quadratic finite elements on a
changing transverse mesh in the branching region. We check the paraxial
approximation against full-wave solutions at selected frequencies.
Separately, we calculate radiation attenuation in a
conducting pipe without an outlet using complex modes that satisfy impedance boundary conditions.
For this example, independently damped perfectly conducting modes
underestimate the high-frequency transmitted power.
For a $70\times25$ mm rectangular chamber with a
$30^\circ$ bend of radius 2.2 m and Gaussian bunches
of rms length $10\,\mu$m, the selected outlet diverts
about 53\% of the total outgoing radiation power into the beam-free channel.
\end{abstract}
\maketitle
\section{Introduction}
\label{sec:intro}
Coherent synchrotron radiation (CSR) in a bending magnet excites modes of
the surrounding vacuum chamber. Stupakov and Kotelnikov developed a
finite-bend mode expansion that relates this excitation to the chamber
geometry and the bunch spectrum~\cite{StupakovKotelnikov2009}. The
radiation affects the electron beam, but it can also carry energy far
beyond the magnet and deposit it along the downstream
pipe~\cite{WarnockBizzozero2016}. For short bunches at high repetition rates, the resulting heat load
can be important in the design of the beam pipe.

We propose a beam-free side outlet near the end of a bend to divert
part of the CSR toward a separate absorber, as illustrated in
Fig.~\ref{fig:baselinegeometry}(a). Optical extraction of CSR
has been used for beam diagnostics~\cite{Hashimoto2005}. Here we seek
to reduce the radiation load in the pipe that carries the electron beam.
We first quantify radiation absorption in a conducting pipe without an
outlet, then study how a perfect-electric-conductor (PEC) junction
partitions radiation between the two guides. The first calculation estimates the distributed heat load from radiation;
the second determines how much radiation can be diverted near the bend.

A useful simplification is possible when the chamber has constant
height. Vertical Fourier expansion reduces the frequency-domain Maxwell
equations to a two-dimensional problem for each vertical harmonic.
Tremaine et al. used horizontal Fourier decomposition for wake fields in
slab-symmetric dielectric structures, including numerical Maxwell
calculations~\cite{Tremaine1997}. Paech applied vertical separation to the
bunch-compressor chamber of the
Free-electron LASer in Hamburg (FLASH)~\cite{Paech2008}.

The paraxial equations of Agoh and
Yokoya~\cite{AgohYokoya2004} and Zhou et al.~\cite{Zhou2012} account for
orbit curvature to leading order. More detailed treatments of the orbit curvature in paraxial equations, including finite-difference (FD) solutions, appear in
Refs.~\cite{Bizzozero2014,WarnockBizzozero2016}. Further developments
include transverse discontinuous Galerkin
methods~\cite{Bizzozero2013} and varying-width
chambers~\cite{Warnock2019}.

The paraxial equation used here retains the geometric factors associated
with the curved beam orbit, rather than expanding all terms to leading
order in curvature. These factors affect the excitation and accumulated
phase of the modes, and hence the radiation reaching the outlet.

We first consider a rectangular bend of constant cross section without
an outlet. For this geometry, we extend the finite-bend modal solution
of Stupakov and Kotelnikov to the paraxial equation used here and
derive an independent conservative FD formulation.

For the downstream conducting pipe, we expand the field in complex modes
satisfying impedance (Robin) boundary conditions. The wall impedance then
changes both the mode profiles and their propagation constants. This
differs from assigning independent attenuation to fixed PEC
modes~\cite{Collin1991}, as in Zhou's straight-guide
calculation~\cite{Zhou2011}. We compare the two approaches
with an independent FD solution.

To calculate extraction, we use quadratic finite elements (FE), denoted
P2, on a transverse mesh that changes along the beam orbit. The mesh
follows the walls and divides into two parts at the splitter. This allows
us to calculate the power in each outgoing guide. We first check the FE
solution against the modal and FD solutions in the bend without an outlet.

We test the paraxial approximation against bidirectional Helmholtz
mode matching in the bend and full-wave Maxwell calculations at the
junction.

As an example, we consider an idealized rectangular chamber with beam
and bend parameters motivated by high-power extreme-ultraviolet
free-electron-laser studies at the High Energy Accelerator Research
Organization (KEK)~\cite{Nakamura2017,Nakamura2023}. We use a Gaussian
bunch of root-mean-square (rms) length $10\,\mu$m and charge 60 pC.
The peak current is 718 A, comparable to the values reported in those
studies. Our Gaussian bunch and rectangular chamber are idealizations,
not a detailed model of the published design.

Section~\ref{sec:fd} describes radiation generation in the bend and
absorption in the downstream pipe without an outlet.
Section~\ref{sec:fe} presents the branching calculation and the dependence
of extraction on outlet geometry. Section~\ref{sec:conclusions}
summarizes the results. Numerical details and the full-wave formulations
are given in Appendices~\ref{app:numerics} and \ref{app:maxwell}.

\section{Bend excitation and downstream absorption}
\label{sec:fd}
\subsection{Geometry and governing equations}
\begin{figure*}[!tbp]
\includegraphics[width=\textwidth]{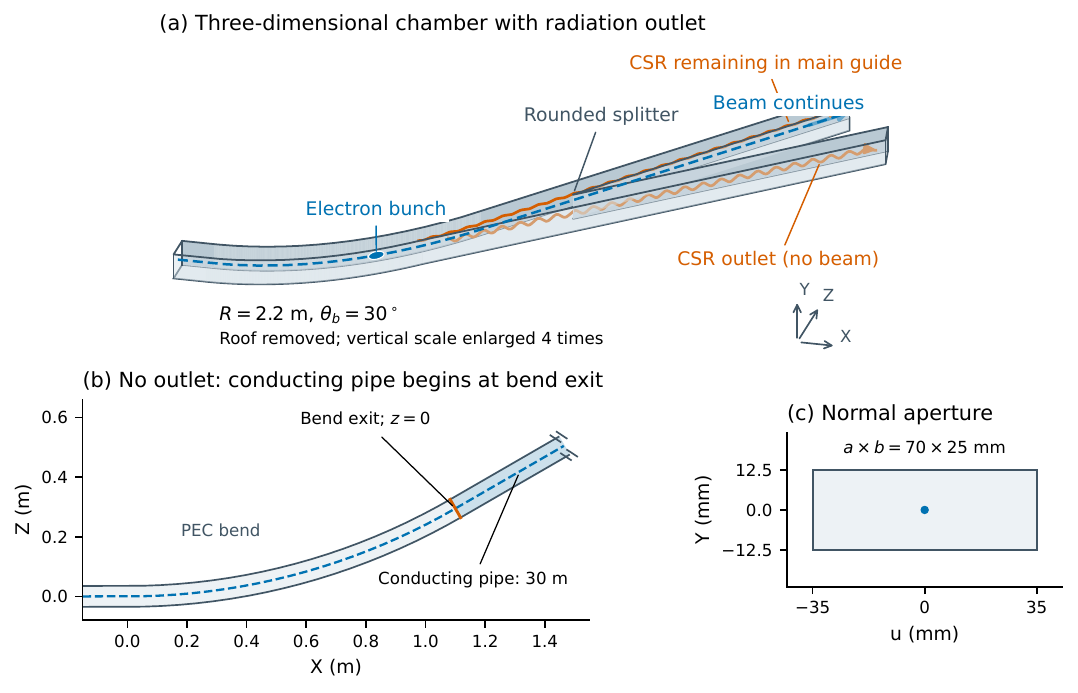}
\caption{Chamber geometry. (a) Beam pipe and outlet with the roof removed
and vertical scale enlarged fourfold. The dashed line is the beam orbit;
the bunch and radiation are schematic. (b) No-outlet reference: a 30 m
conducting pipe starts at $z=s-L_b=0$; only its entrance is shown.
(c) Rectangular aperture $a\times b=70\times25$ mm.}
\label{fig:baselinegeometry}
\end{figure*}

We first calculate radiation in the rectangular bend without an outlet.
Its exit field provides the initial condition for the downstream
conducting pipe. Figure~\ref{fig:baselinegeometry} shows the chamber
and reference orbit.
The rectangular aperture has horizontal width $a$ and vertical height $b$.
The reference orbit has radius $R$, bend angle $\theta_b$, and length
$L_b=R\theta_b$, with $\theta_b$ in radians. The longitudinal coordinate
$s$ is distance along this orbit, measured from the magnet entrance;
the magnet exit is at $s=L_b$. The transverse coordinate $u$ is the
signed horizontal distance from the orbit along its local normal,
positive away from the bend center. The vertical coordinate $Y$ is
measured from the chamber midplane. The beam follows $u=Y=0$;
the aperture spans $|u|\leq a/2$ and $|Y|\leq b/2$.

The 70 mm width follows the horizontal aperture
scale used in the second-arc transport study of
Ref.~\cite{Nakamura2017}; the height of 25 mm is a rounded value based on the Cornell
study~\cite{CornellPDDR}. We use these dimensions for a constant
rectangular cross section. The parameters are collected in Table~\ref{tab:benchmark}.

The fixed Cartesian coordinates $(X,Z)$ describe the horizontal plane:
their origin is the magnet entrance, $X$ follows the entrance tangent,
and $Z$ points toward the bend center. In the magnet,
$0\leq s\leq L_b$, the position $\bm r=(X,Z)$ is
\begin{equation}
 \begin{split}
 \bm r(s,u)={}&R\bigl(\sin(s/R),1-\cos(s/R)\bigr)\\
             &+u\bigl(\sin(s/R),-\cos(s/R)\bigr).
 \end{split}
 \label{eq:orbit}
\end{equation}
The geometric factor $h=1+\kappa u$ converts a longitudinal
coordinate increment $\dd s$ into the physical distance $h\dd s$
at fixed $u$.
The orbit curvature $\kappa$ is $1/R$ in the magnet and zero in its
tangent continuations. 

\begin{table}[tb]
\caption{Parameters used in the calculations. Beam and bend parameters are
motivated by Refs.~\cite{Nakamura2017,Nakamura2023}; the bunch shape,
chamber geometry, and wall conductivity are idealized.}
\label{tab:benchmark}
\begin{ruledtabular}\begin{tabular}{ll}
Quantity&Value\\
Gaussian rms length $\sigma_s$&$10\,\mu$m\\
Bunch charge $Q$; repetition $f_{\rm rep}$&60 pC; 162.5 MHz\\
Mean current $Qf_{\rm rep}$&9.75 mA\\
Radius $R$; bend angle $\theta_b$&2.2 m; $30^\circ$\\
Bend length $L_b$&1.152 m\\
Normal aperture $a\times b$&$70\times25$ mm\\
Downstream conductivity $\sigma_c$&$3.5\times10^7$ S/m\\
Conducting interval from bend exit&30 m\\
\end{tabular}\end{ruledtabular}
\end{table}

We assume that the bunch moves at the speed of light $c$ without changing
shape. The bend and outlet have PEC walls. We use time dependence
$e^{-\ii\omega t}$, where $t$ is time, $f$ is frequency,
$\omega=2\pi f$ is angular frequency, $\ii^2=-1$, and $k=\omega/c$
is the vacuum wave number. The vacuum impedance is
$\Z=\sqrt{\mu_0/\epsilon_0}$, where $\mu_0$ and $\epsilon_0$ are the
vacuum permeability and permittivity. We denote real and imaginary
parts by $\operatorname{Re}$ and $\operatorname{Im}$, respectively, and
complex conjugation by a superscript $*$.

The constant chamber height permits separation of the vertical coordinate.
Let $m$ label a vertical harmonic and put $\alpha_m=m\pi/b$. The Fourier
coefficients $F_m^E$ and $F_m^H$ represent the vertical
electric and magnetic fields:
\begin{align}
 E_Y(s,u,Y)&=\sum_m F_m^E(s,u)\cos[\alpha_m(Y+b/2)],\nonumber\\
 H_Y(s,u,Y)&=\sum_m F_m^H(s,u)\sin[\alpha_m(Y+b/2)].
 \label{eq:verticalbasis}
\end{align}
Superscripts E and H identify the two scalar families. Charge density
and horizontal current use the sine basis. For a line source on the
midplane $Y=0$, only odd vertical harmonics, $m=1,3,\ldots$, are excited.
We suppress the family and mode labels in equations common to both.

We write the total scalar field as
$F(s,u)=e^{\ii ks}[G(u)+v(s,u)]$.
Here $G$ is the known beam-field envelope in a straight PEC guide
of the same aperture, for the same source and vertical harmonic.
It is independent of $s$ after factoring out $e^{\ii ks}$.
The remainder $v$ is the radiation envelope.
Let $I_\omega$ denote the complex peak amplitude of the beam-current
harmonic at angular frequency $\omega$. The sine expansion of the
midplane source $\delta(Y)$ has coefficients $(2/b)\sin(m\pi/2)$.
For odd $m$, we include this source factor in
$C_m=I_\omega(-1)^{(m-1)/2}/(c\epsilon_0b)$ and set $d=a/2$.
The straight-guide particular fields are
\begin{align}
 G^E&=-C_m\frac{\sinh[\alpha_m(d-|u|)]}{\cosh(\alpha_m d)},\nonumber\\
 G^H&=\frac{C_m}{\Z}\operatorname{sgn}(u)
              \frac{\cosh[\alpha_m(d-|u|)]}{\cosh(\alpha_m d)}.
 \label{eq:mrparticular}
\end{align}
Here $\operatorname{sgn}(u)$ is the sign of $u$. The electric field
$G^E$ is continuous, while $G^H$ jumps by $2C_m/\Z$ at the beam.
Subtracting these known fields removes the delta-function source terms,
so the remainder can be computed without representing them on the mesh.

The paraxial equation used here follows from the scalar
Helmholtz equation after neglecting the second longitudinal derivative
$h^{-1}\partial_s^2v$ of the envelope:
\begin{align}
 \frac{2\ii k}{h}\partial_s v+\mathcal L_m v&=r_G,\nonumber\\
 \mathcal L_m&=\partial_u(h\partial_u)
                +k^2(h-h^{-1})-\alpha_m^2h,\nonumber\\
 r_G&=-\kappa G'_{\rm reg}-k^2(h-h^{-1})G.
 \label{eq:metricmain}
\end{align}
Here $\partial_s$ and $\partial_u$ denote partial derivatives,
$\mathcal L_m$ is the transverse operator, and $r_G$ is the source left
after subtracting $G$. The subscript ``reg'' denotes the ordinary
$u$ derivative away from the beam:
$(G^E)'_{\rm reg}=\alpha_m\Z G^H$ and
$(G^H)'_{\rm reg}=\alpha_m G^E/\Z$.
The delta-function contributions have already canceled; their treatment
is given in Appendix~\ref{sec:fenumerics}.
In a rectangular bend without an outlet, the PEC boundary conditions
are $v^E(\pm d)=0$ and $\partial_uv^H(\pm d)=0$. We impose
$v(0,u)=0$ at the entrance, corresponding to a bunch entering from a
straight guide with its equilibrium beam field and no incident radiation.

Agoh and Yokoya use the leading paraxial equation for the transverse
electric field~\cite{AgohYokoya2004}. After vertical Fourier separation,
their beam-field-subtracted Eq.~(29), written in our notation, is
\begin{equation}
 2\ii k\partial_s v+
 \left(\partial_u^2-\alpha_m^2+2k^2\kappa u\right)v
 =-2k^2\kappa uG.
 \label{eq:leadingparaxial}
\end{equation}
For the vertical electric family this identification is direct. The
horizontal electric component $E_u$ gives the same leading equation
for the magnetic scalar through $E_u=\Z H_Y$ to paraxial order. The term
$2k^2\kappa u$ accounts for orbit curvature in this approximation.

We instead solve Eq.~\eqref{eq:metricmain}, retaining $h^{-1}$ in the
longitudinal derivative and $h$ in the transverse operator. In a
constant-curvature segment,
$\partial_u(h\partial_u v)=h\partial_u^2v+\kappa\partial_uv$.
The curvature coefficient is $k^2(h-h^{-1})$, and the source includes
$-\kappa G'_{\rm reg}$. Replacing the derivative weights by unity,
using $h-h^{-1}\simeq2\kappa u$, and dropping this derivative of $G$
recovers Eq.~\eqref{eq:leadingparaxial}. The two equations thus differ in the derivative terms and the source,
not just in the curvature coefficient.

Although $|u/R|\leq0.016$ across the aperture, these terms affect the
excitation and accumulated phase of the modes. As shown in
Sec.~\ref{sec:bendcontrol}, Eq.~\eqref{eq:metricmain} gives closer
agreement with the full-wave solution. We use it to calculate the bend
field entering either the conducting pipe or the outlet region.

Both equations neglect the second longitudinal derivative of the
envelope and describe only forward propagation. Keeping the geometric
terms does not give a complete higher-order approximation, so comparison
with a full-wave solution is still needed. Equation~\eqref{eq:metricmain}
has been used in Refs.~\cite{Bizzozero2014,WarnockBizzozero2016}; here we
solve it by modal, FD, and FE methods. Appendix~\ref{app:numerics}
discusses the order of the neglected terms.

\subsection{Modal and finite-difference solutions}
\label{sec:bendmethod}
We extend the finite-bend mode-expansion method of Stupakov and
Kotelnikov~\cite{StupakovKotelnikov2009} from
Eq.~\eqref{eq:leadingparaxial} to Eq.~\eqref{eq:metricmain}.
The transverse eigenfunctions are then orthogonal with weight $h^{-1}$,
and the curvature-dependent factors enter the operator and source.
For a bend of constant curvature, $h$ is independent of $s$.
We then expand the radiation in transverse eigenfunctions. For a fixed vertical harmonic and scalar
family, the horizontal modes $\psi_\nu$ satisfy
\begin{equation}
 \mathcal L_m\psi_\nu=\lambda_\nu h^{-1}\psi_\nu,
 \qquad \int_{-d}^{d}\frac{\psi_\mu\psi_\nu}{h}\dd u=\delta_{\mu\nu}.
 \label{eq:weightedeigenmain}
\end{equation}
The indices $\mu,\nu$ label horizontal modes, $\lambda_\nu$ are their
eigenvalues, and $\delta_{\mu\nu}$ is the Kronecker delta. The PEC
conditions eliminate the boundary term in integration by parts.
Since $h>0$, $h\mathcal L_m$ is self-adjoint in the $h^{-1}$-weighted
inner product, and the eigenfunctions and eigenvalues can be chosen real.

The expansion $v=\sum_\nu c_\nu(s)\psi_\nu$ defines the modal amplitudes
$c_\nu$. Projection of Eq.~\eqref{eq:metricmain} gives independent
ordinary differential equations,
\begin{equation}
 2\ii k c_\nu'+\lambda_\nu c_\nu=b_\nu,
 \qquad b_\nu=\int_{-d}^{d}\psi_\nu r_G\dd u,
 \label{eq:mrexcitation}
\end{equation}
where $c_\nu'=\dd c_\nu/\dd s$ and $b_\nu$ is the source projection.
For a constant-curvature segment of length $L$,
\begin{equation}
 c_\nu(L)=c_\nu(0)e^{\ii\lambda_\nu L/(2k)}
       +\frac{b_\nu}{\lambda_\nu}
                         [1-e^{\ii\lambda_\nu L/(2k)}].
 \label{eq:mrresponse}
\end{equation}
The bend-exit solution follows by setting $L=L_b$ and $c_\nu(0)=0$.
The source term tends to $-\ii Lb_\nu/(2k)$ as $\lambda_\nu\to0$;
we use this limit to avoid cancellation. 

In the leading-order formulation of Stupakov and Kotelnikov,
the eigenfunctions for a rectangular chamber can be expressed
in terms of Airy functions.
Here we determine the eigenfunctions numerically by expanding
them in real orthonormal straight-guide functions.
For the two scalar families, these are
\begin{align}
 \phi_n^E(u)&=\sqrt{2/a}\sin[n\pi(u+d)/a],\quad n\geq1,\nonumber\\
 \phi_n^H(u)&=\sqrt{(2-\delta_{n0})/a}\cos[n\pi(u+d)/a],\quad n\geq0.
 \label{eq:portbasis}
\end{align}
Here $n$ is the horizontal mode index; the zero-index magnetic
function is constant.
We represent each eigenfunction as
$\psi_\nu(u)=\sum_i(q_\nu)_i\phi_i(u)$,
where $(q_\nu)_i$ are the expansion coefficients.
Suppressing the family label, we define
$W_{ij}=\int\phi_i\phi_j/h\dd u$ and
$K_{ij}=\int\phi_i\mathcal L_m\phi_j\dd u$,
with the derivative term in $K_{ij}$ integrated by parts.
The coefficients $\bm q_\nu$ and eigenvalues $\lambda_\nu$
are determined from the generalized eigenvalue problem
\begin{equation}
\begin{aligned}
K\bm q_\nu &= \lambda_\nu W\bm q_\nu,\\
\bm q_\mu^T W\bm q_\nu &= \delta_{\mu\nu},
\end{aligned}
\end{equation}
where the second equation specifies the normalization
and $T$ denotes transpose.
All integrals extend over $-d\leq u\leq d$.
We split the quadrature at the beam position $u=0$ and check convergence
by increasing the number of basis functions.

As an independent check, we discretize the transverse operator in
Eq.~\eqref{eq:metricmain} using differences of the flux
$h\partial_u v$ at cell faces.
In this discretization, $W$, $K$, and $\bm b$ denote the nodal mass,
stiffness, and load. The nodes are $u_j=-d+j\Delta u$, $j=0,\ldots,J$,
where $J$ is the number of intervals and $\Delta u=2d/J$ is their
spacing. The trapezoidal quadrature weights are
$\omega_j=\Delta u$ at interior nodes and
$\omega_j=\Delta u/2$ at the retained Neumann endpoints. Electric Dirichlet endpoints
are eliminated. With $h_j=1+u_j/R$ and
$h_{j+1/2}=(h_j+h_{j+1})/2$, we obtain
\begin{align}
 W_{jj}&=\omega_j/h_j,\nonumber\\
 K&=-D^T\operatorname{diag}(h_{j+1/2}/\Delta u)D\nonumber\\
       &\quad+\operatorname{diag}\{\omega_j[k^2(h_j-h_j^{-1})
-\alpha_m^2h_j]\},\nonumber\\
 b_j&=\omega_j r_G(u_j).
 \label{eq:mrfd}
\end{align}
Here $D$ is the first-difference matrix,
$(D\bm v)_j=v_{j+1}-v_j$, acting on the nodal vector $\bm v$;
$\operatorname{diag}$ forms a diagonal matrix from its entries. At
$u=0$, the nodal load uses the mean of the two one-sided regular values.
The magnetic boundary flux is zero. The resulting system is
$2\ii kW\bm v'+K\bm v=\bm b$, with $\bm v'=\dd\bm v/\dd s$.
The mass matrix $W$ is positive diagonal, and $K$ is real symmetric.
Related FD approaches are described in
Refs.~\cite{Bizzozero2014,WarnockBizzozero2016}.

The transformation $\bm y=W^{1/2}\bm v$ gives the real symmetric
tridiagonal matrix
$B=W^{-1/2}KW^{-1/2}$.
Diagonalizing $B$ and using Eq.~\eqref{eq:mrresponse} gives exact
longitudinal propagation of the spatially discretized system. We can
therefore study transverse convergence without a longitudinal stepping
error. On the largest grids we retain only part of the FD eigenbasis
and check convergence with respect to its size. These are eigenvectors
of the FD matrix, not the continuum modes introduced above.
For a longitudinal-stepping check, we also propagate the same FD system
with Crank--Nicolson (CN)~\cite{CrankNicolson1947}. Comparing this result
with exact propagation isolates the longitudinal discretization error.
Without a source, both methods conserve $\bm v^\dagger W\bm v$, where
$\dagger$ denotes conjugate transpose, but their phases can differ.
Appendix~\ref{sec:wallnumerics} gives the propagation factors and checks.

\subsection{Bend excitation and the role of geometry}
\label{sec:bendcontrol}
Agreement between numerical methods does not test the paraxial
approximation itself. For this purpose we solve the Helmholtz equation,
retaining both propagation directions. We write the radiation field as
$U=e^{\ii ks}v$, restoring the phase factor removed from the envelope.
Thus, for each scalar family and vertical mode, we write
$F=e^{\ii ks}G+U$.
In a constant-curvature segment, $U$ obeys
\begin{equation}
 h^{-1}U_{ss}+\partial_u(hU_u)+h(k^2-\alpha_m^2)U
 =r_Ge^{\ii ks}.
 \label{eq:helmholtzcontrol}
\end{equation}
Substituting $U=e^{\ii ks}v$ into this equation and neglecting
the second longitudinal derivative $v_{ss}$ recovers
the paraxial Eq.~\eqref{eq:metricmain}.
Here $r_G$ is the same residual source as in Eq.~\eqref{eq:metricmain},
and subscripts denote partial derivatives. The side walls have the same
PEC conditions. Straight guides extend from both ends of the bend, with
outgoing or decaying radiation and no incident radiation. At each
bend--straight interface, the total field and its physical normal
derivative are continuous. This derivative is $h^{-1}\partial_sF$ on
the curved side and $\partial_sF$ on the straight side, with the same
orientation. These conditions determine $U$ including reflected waves.
Appendix~\ref{app:maxwell} derives the modal matching system used to
solve this boundary-value problem.

We compare radiation on the straight side of the magnet exit, $s=L_b$.
All numerical field comparisons use $I_\omega=1$ A.
Using the real functions in Eq.~\eqref{eq:portbasis}, define
\begin{align}
 a_{nm}^j&=e^{\ii kL_b}\int_{-d}^{d}\phi_n^j(u)v_m^j(L_b,u)\dd u,
 \nonumber\\
 a_{{\rm Helm},nm}^j&=\int_{-d}^{d}\phi_n^j(u)U_m^j(L_b,u)\dd u.
 \label{eq:outgoingamplitudes}
\end{align}
Here $j=E,H$ labels the scalar family, $n$ the horizontal mode, and
$m$ the vertical mode. The first line projects the paraxial radiation;
the second projects the Helmholtz radiation after subtraction of the
same straight-guide beam field. Thus both amplitudes have the same
normalization and carrier phase. The physical outgoing
power of vertical mode $m$ is
\begin{align}
 P_m&=\frac b4\sum_{\beta_{nm}>0}
 \frac{k\beta_{nm}}{k^2-\alpha_m^2}
 \left(\frac{|a^E_{nm}|^2}{\Z}+\Z|a^H_{nm}|^2\right),\nonumber\\
 \beta_{nm}&=\sqrt{k^2-\alpha_m^2-(n\pi/a)^2}.
 \label{eq:portflux}
\end{align}
Here $\beta_{nm}$ is the longitudinal wave number; the sum includes only
propagating modes, for which it is real and positive. The E family starts at
$n=1$ and H includes $n=0$; missing coefficients
in the combined notation are zero. The factor $b/4$ combines the integral $b/2$ of the squared vertical sine or cosine function over the chamber height
with the factor $1/2$ from time averaging the harmonic fields.

Let $\bm a$ and $\bm a_{\rm Helm}$ contain the propagating-mode
coefficients from the paraxial and Helmholtz solutions, respectively,
including both E and H families.
Both fields use the same geometry, line source, frequency, vertical
index, and harmonic current; no phase or amplitude is fitted.
Equation~\eqref{eq:portflux} defines the power
norm $\|\bm a\|_P^2=P_m$. The Helmholtz radiation power is
$P_{\rm Helm}=\|\bm a_{\rm Helm}\|_P^2$. We compare them through
\begin{equation}
 \delta_P=100\frac{P_m-P_{\rm Helm}}{P_{\rm Helm}},\qquad
 \epsilon_F=100\frac{\|\bm a-\bm a_{\rm Helm}\|_P}
                         {\|\bm a_{\rm Helm}\|_P}.
 \label{eq:benderrors}
\end{equation}
The signed difference $\delta_P$ compares radiation powers,
whereas $\epsilon_F$ measures the difference between complex
fields, including phase and transverse profile.
Only propagating modes contribute to these measures.

At 0.24 THz ($m=1$), the leading-order paraxial Eq.~\eqref{eq:leadingparaxial}
differs from the bidirectional Helmholtz result by about 9\% in
radiation power; using Eq.~\eqref{eq:metricmain} reduces this
difference to 0.05\%. The complex-field discrepancy remains
about 2\%, showing that agreement in power does not imply equally
accurate phases or transverse field profiles.

\begin{figure*}[!tbp]
\includegraphics[width=\textwidth]{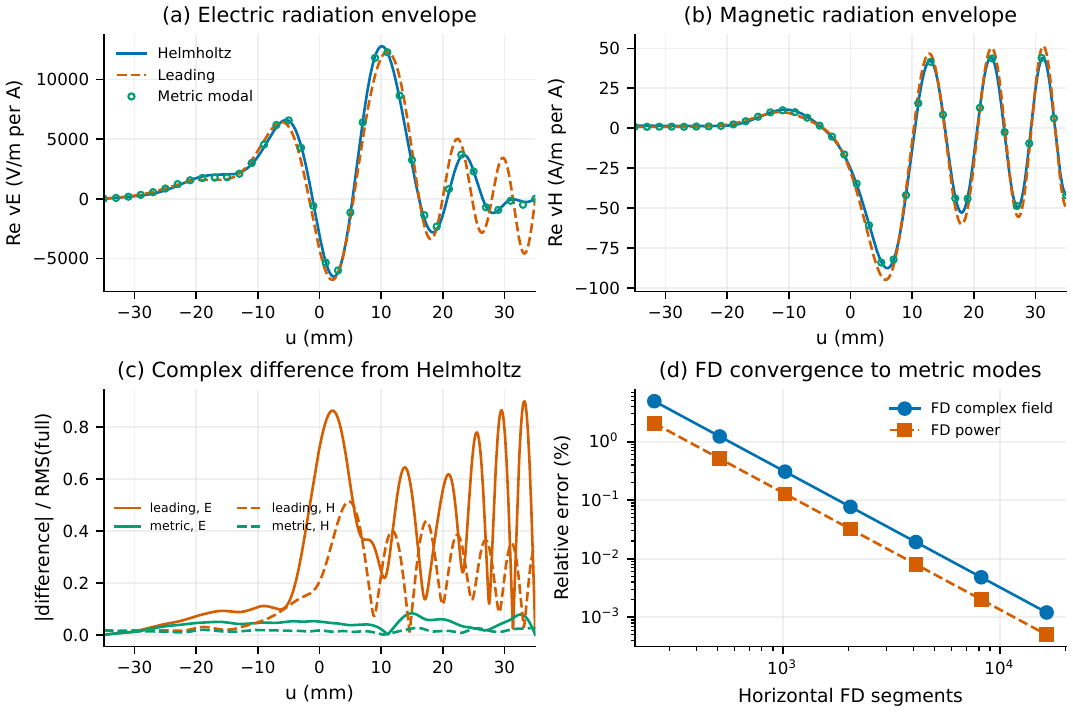}
\caption{Bend-exit radiation at 0.24 THz, $m=1$, $I_\omega=1$ A.
(a,b) Real parts of the envelopes $v^E$ and $v^H$. ``Leading'' and
``Metric modal'' denote Eqs.~\eqref{eq:leadingparaxial} and
\eqref{eq:metricmain}; Helmholtz uses the same line source.
(c) Pointwise complex-field differences divided by the corresponding
Helmholtz field's spatial rms amplitude. (d) FD convergence to the modal
solution of Eq.~\eqref{eq:metricmain}, measured by $\epsilon_F$ and
$|\delta_P|$ [Eq.~\eqref{eq:benderrors}].}
\label{fig:bendcontrol}
\end{figure*}

Figure~\ref{fig:bendcontrol} illustrates these results
and distinguishes errors of the paraxial approximation
from numerical discretization errors.
Panels (a--c) compare the two paraxial equations with the
Helmholtz reference. Panel (d) shows convergence of the
conservative FD calculation to the modal solution of
Eq.~\eqref{eq:metricmain}.

Table~\ref{tab:bendfrequency} summarizes the frequency
dependence of the discrepancies for the lowest vertical
harmonic ($m=1$). We then examine how these discrepancies
change when all retained vertical harmonics are included.

\begin{table}[tb]
\caption{Comparison of the paraxial solution of Eq.~\eqref{eq:metricmain}
with the full-wave Helmholtz solution at the bend exit for $m=1$.
Both use the same line source. The power difference $\delta_P$ and
complex-field difference $\epsilon_F$ are defined in
Eq.~\eqref{eq:benderrors} and reported in percent.}
\label{tab:bendfrequency}
\begin{ruledtabular}\begin{tabular}{rrr}
$f$ (THz)&$\delta_P$ (\%)&$\epsilon_F$ (\%)\\
0.12&+0.05&1.2\\
0.24&+0.05&2.0\\
0.50&-0.01&5.4\\
1.00&-0.23&9.1\\
3.00&-0.81&26.1\\
6.00&-0.38&31.9\\
\end{tabular}\end{ruledtabular}
\end{table}

Summing all retained odd vertical harmonics gives a similar
improvement in power. At 0.24 THz ($m\leq15$), the discrepancy
relative to Helmholtz decreases from about 8\% with
Eq.~\eqref{eq:leadingparaxial} to 0.05\% with
Eq.~\eqref{eq:metricmain}; the corresponding complex-field
discrepancy is about 2\%.

At higher frequencies, the distinction between power
and field agreement becomes more pronounced.
At 3 THz, with odd $m\leq37$, the solution of
Eq.~\eqref{eq:metricmain} agrees with Helmholtz to within 0.2\%
in total radiation power, but the complex-field
discrepancy reaches about 12\%.
The 26\% discrepancy for $m=1$ alone is therefore not representative
of the full field: higher vertical harmonics carry a significant part
of the power at this frequency.
The remaining field discrepancy persists under
horizontal basis refinement, indicating a limitation
of the one-way approximation rather than insufficient
numerical resolution.

\begin{figure*}[!tbp]
\includegraphics[width=\textwidth]{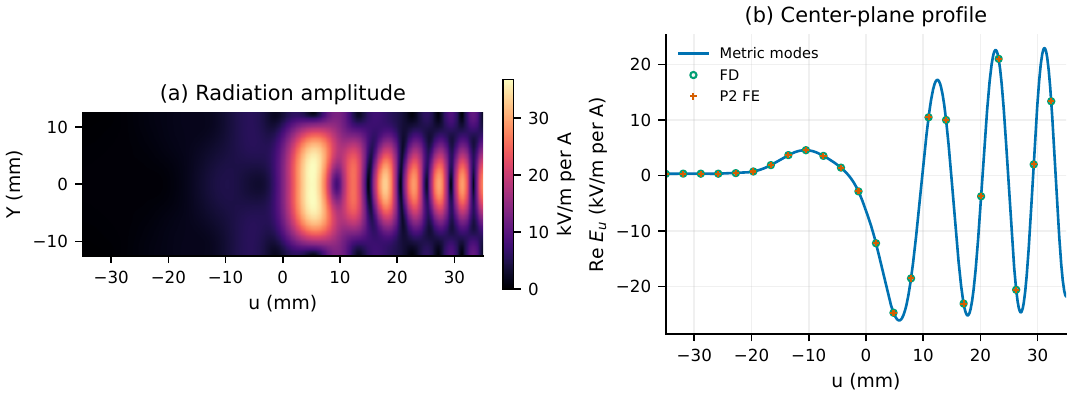}
\caption{Radiation at the bend exit for 0.24 THz, $m=1,3,\ldots,15$,
and $I_\omega=1$ A. (a) Modal amplitude $\sqrt{|E_u|^2+|E_Y|^2}$
from Eq.~\eqref{eq:metricmain}. (b) $\operatorname{Re}E_u$ at $Y=0$
from modal, FD, and P2 solutions. The beam field is subtracted, with
$E_u=\Z H_Y$ to paraxial order.
}
\label{fig:fullfield}
\end{figure*}

Figure~\ref{fig:fullfield} shows the reconstructed
radiation field at 0.24 THz, including all retained
vertical harmonics.
The modal, conservative FD, and P2 solutions of
Eq.~\eqref{eq:metricmain} agree at the field level to better than
0.01\%. This independently checks the numerical implementations
of the same paraxial equation.
The P2 formulation is subsequently used for the
branching geometry in Sec.~\ref{sec:fe}.

\subsection{Bunch spectrum and wall losses}
\label{sec:absorption}

The bend calculation gives the radiation field at each frequency.
We first combine these fields with the bunch spectrum to obtain
the average radiation power. We then follow this radiation through
the conducting pipe downstream of the bend.

At a straight section, we use the paraxial expression for the
radiation power per squared harmonic current,
\begin{equation}
 \mathcal P(f,s)=\frac{b}{4|I_\omega|^2}\sum_{m\ \mathrm{odd}}
 \int\left(|v_m^E|^2/\Z+\Z|v_m^H|^2\right)\dd u.
 \label{eq:paraxialpower}
\end{equation}
The integral extends over the guide width. Since the fields are
proportional to $I_\omega$, division by $|I_\omega|^2$ makes
$\mathcal P$, in W/A$^2$, independent of the chosen current amplitude.
Compared with the outgoing-mode flux in Eq.~\eqref{eq:portflux},
this expression omits the longitudinal-wave-number factors and
includes all retained transverse components.
The two expressions agree in the forward paraxial limit.
We use Eq.~\eqref{eq:paraxialpower} for the broadband calculation.

For a periodic train of Gaussian bunches with charge $Q$,
repetition rate $f_{\rm rep}$, and rms length $\sigma_s$,
the positive-frequency harmonics have frequencies and peak currents
\begin{equation*}
\begin{aligned}
 f_\ell&=\ell f_{\rm rep},\qquad \ell=1,2,\ldots,\\
 I_\ell&=2Qf_{\rm rep}
 \exp\left[-\frac12\left(\frac{2\pi f_\ell\sigma_s}{c}\right)^2\right].
\end{aligned}
\end{equation*}
Each harmonic contributes $|I_\ell|^2\mathcal P(f_\ell,s)$
to the average power. Replacing the sum over harmonics,
spaced by $f_{\rm rep}$, with an integral gives
\begin{align}
 S(f,s)&=\frac{\dd P}{\dd f}
 =4Q^2 f_{\rm rep}\,e^{-(2\pi f\sigma_s/c)^2}\mathcal P(f,s),\nonumber\\
 P(s)&=\int_0^\infty S(f,s)\dd f.
 \label{eq:bunchpower}
\end{align}
Here $S$ is the spectral power density in W/Hz.
For a fixed bunch shape, the power is proportional to
$Q I_{\rm av}$, where $I_{\rm av}=Qf_{\rm rep}$.

We now consider the pipe without an outlet.
The conducting section begins at the magnet exit, and we measure
distance along it by $z=s-L_b$.
The radiation field calculated at the bend exit provides the
initial condition at $z=0$.
Let $\bm A=(A_u,A_Y)$ denote its transverse electric envelope
divided by $I_\omega$.
At the entrance, $A_Y$ is reconstructed from
$v_m^E(L_b,u)/I_\omega$ using the vertical cosine functions
in Eq.~\eqref{eq:verticalbasis}.
Similarly, $A_u$ is reconstructed from
$\Z v_m^H(L_b,u)/I_\omega$ using the sine functions.

In the straight pipe, each component $j=u,Y$ obeys a paraxial
propagation equation. Finite wall conductivity enters through
the boundary conditions:
\begin{align}
 \partial_z A_j&=\frac{\ii}{2k}\nabla_\perp^2A_j,\nonumber\\
 \partial_n A_n&=\ii k\chi A_n,\qquad A_t=0,\nonumber\\
 \chi&=(1-\ii)\sqrt{\frac{k}{2\Z\sigma_c}}.
 \label{eq:robin}
\end{align}
Here $\nabla_\perp^2=\partial_u^2+\partial_Y^2$,
$\partial_n$ is the outward normal derivative, and $A_n$ and
$A_t$ are the field components normal and tangential to a wall.
The condition on $A_n$ relates the field to its normal derivative
and is a Robin boundary condition.
The dimensionless parameter $\chi$ describes the wall response.
We use the component-wise paraxial surface-impedance model
in the normal skin-effect regime, with a frequency-independent
conductivity $\sigma_c$.
For a particular metal at THz frequencies, a more detailed
surface-impedance model may be needed%
~\cite{BaneStupakov2015,BaneStupakovErratum2016}.

The remaining radiation power per squared current and its loss
per unit length are
\begin{equation}
 \begin{aligned}
 p(f,z)&=\frac1{2\Z}\int_{\mathcal A}|\bm A|^2\dd A,\\
 \partial_z p(f,z)&=
 -\frac{\operatorname{Re}\chi}{2\Z}
 \sum_j\int_{\Gamma_j}|A_j|^2\dd\ell.
 \end{aligned}
 \label{eq:wallbalance}
\end{equation}
The first integral is over the cross-sectional area $\mathcal A$;
the second is over the wall pair $\Gamma_j$ normal to component $j$.
Thus the decrease in radiation power equals the power absorbed
by the walls.
At the entrance, $p(f,0)=\mathcal P(f,L_b)$.
At any downstream position, the spectrum $S(f,z)$ and average
power $P(z)$ follow from Eq.~\eqref{eq:bunchpower}, with
$p(f,z)$ in place of $\mathcal P(f,s)$.

This calculation propagates only the radiation obtained after
subtracting the straight-guide beam field.
It does not include the beam's resistive-wall field or its
interference with the radiation.
The calculated deposition is therefore not the total heat load.
Because the losses are quadratic in the field, the total load
cannot in general be obtained by simply adding this result
to the beam-field loss.

To propagate the radiation, we expand it in modes that satisfy
the impedance boundary conditions.
The wall impedance then changes both the transverse profiles
and the propagation constants.
This differs from the approximation of independently damped
PEC modes~\cite{Collin1991}, used by Zhou~\cite{Zhou2011},
in which the profiles are kept fixed and each mode is assigned
an attenuation coefficient.

The rectangular cross section allows separation of the two
transverse coordinates.
For each electric-field component, we call $x$ the coordinate
between the walls where that component is normal and satisfies
the Robin condition. The other coordinate is $y$, along which
the field vanishes at the walls.
For $A_u$, we therefore use $(x,y)=(u,Y)$ and $D=a$;
for $A_Y$, we use $(x,y)=(Y,u)$ and $D=b$.
The width in $x$ is $D$, and the width in $y$ is $ab/D$.

In these coordinates, we write
\begin{equation*}
 A_j(x,y,z)=\sum_{\ell,r}
 d_{\ell r}(z)\phi_\ell(x)\zeta_r(y).
\end{equation*}
Here $\ell$ labels the Robin modes in $x$, and $r$ labels
the sine modes in $y$. The component index $j$ is suppressed
on the coefficients and mode functions.
The normalized sine functions are
\begin{equation*}
\begin{aligned}
 \zeta_r(y)&=\sqrt{\frac{2D}{ab}}\,
 \sin\left[k_r\left(y+\frac{ab}{2D}\right)\right],\\
 k_r&=\frac{r\pi D}{ab},\qquad r=1,2,\ldots.
\end{aligned}
\end{equation*}

The Robin profiles $\phi_\ell(x)$ on $-D/2\leq x\leq D/2$
are found from
\begin{equation}
\begin{aligned}
 \phi''+q^2\phi&=0,\\
 \phi'(D/2)&=\ii k\chi\phi(D/2),\\
 -\phi'(-D/2)&=\ii k\chi\phi(-D/2),
\end{aligned}
\end{equation}
where primes denote differentiation with respect to $x$.
The even and odd solutions are proportional to $\cos(qx)$
and $\sin(qx)$, respectively. Substituting these functions
into the boundary conditions gives
\begin{equation}
 -q\tan(qD/2)=\ii k\chi,\qquad
 q\cot(qD/2)=\ii k\chi,
 \label{eq:robinroots}
\end{equation}
for the even and odd modes.
We obtain the complex roots $q_\ell$ by starting with the
Neumann eigenvalues at zero impedance and gradually increasing
the impedance to its required value.

The initial amplitudes are found by projecting the bend-exit
field onto these modes.
For distinct eigenvalues, Green's identity gives orthogonality
without complex conjugation.
Away from degeneracies, we normalize the Robin functions so that
$\int_{-D/2}^{D/2}\phi_\ell\phi_{\ell'}\,\dd x
=\delta_{\ell\ell'}$.
The projection is then
\begin{equation}
 d_{\ell r}(0)=\int_{\mathcal A}
       \phi_\ell(x)\zeta_r(y)A_j(x,y,0)\dd A.
 \label{eq:robinprojection}
\end{equation}
We evaluate these integrals using analytic overlaps between
the Robin functions and the straight-guide sine and cosine
functions of the bend-exit field.
No phase or amplitude adjustment is made.

Once the initial amplitudes are known, each mode propagates
independently:
\begin{equation*}
\begin{aligned}
 d_{\ell r}(z)&=d_{\ell r}(0)e^{\lambda_{\ell r}z},\\
 \lambda_{\ell r}&=-\frac{\ii(q_\ell^2+k_r^2)}{2k}.
\end{aligned}
\end{equation*}
The real and imaginary parts of $\lambda_{\ell r}$ determine
the attenuation and phase advance along the pipe.

Although the amplitudes propagate independently, their powers
cannot in general be added independently.
The power involves products $\phi_\ell^*\phi_{\ell'}$,
for which the complex Robin modes are not orthogonal.
We therefore retain the interference terms through an overlap
matrix $\mathsf G$.
With $\bm d_r$ collecting the amplitudes at fixed $r$,
the power of component $j$ is
\begin{align}
 \mathsf G_{\ell\ell'}&=
 \int_{-D/2}^{D/2}\phi_\ell^*\phi_{\ell'}\,\dd x,\nonumber\\
 p_j(z)&=\frac1{2\Z}\sum_r
             \bm d_r(z)^\dagger\mathsf G\bm d_r(z).
 \label{eq:robingram}
\end{align}
The matrix $\mathsf G$ is the Hermitian Gram matrix,
and the total normalized power is $p=p_u+p_Y$.
We check its decrease with $z$ against the wall-loss
integral in Eq.~\eqref{eq:wallbalance}.

As an independent check, we solve the same propagation equation
and boundary conditions with finite differences.
The discrete modes are propagated exactly along $z$,
removing longitudinal stepping error from the comparison.
Appendix~\ref{app:numerics} gives the wall stencil and
the discrete eigenproblem.

For the comparison labeled ``damped PEC'', the amplitude attenuation
coefficient $\alpha_{nm}^{\rm PEC}$, in inverse meters, is the standard
waveguide result~\cite{Collin1991} used in Eq.~(2.57) of Zhou's
thesis~\cite{Zhou2011}:
\begin{equation}
 \alpha_{nm}^{\rm PEC}=
 \frac{2\operatorname{Re}\chi}{ab}\frac{n^2b^3+m^2a^3}{n^2b^2+m^2a^2}.
 \label{eq:prescribedattenuation}
\end{equation}
Each PEC amplitude acquires the factor
$\exp[-\alpha_{nm}^{\rm PEC}z-\ii k_{c,nm}^2z/(2k)]$, where
$k_{c,nm}^2=(n\pi/a)^2+(m\pi/b)^2$. The permitted indices are
$n\geq0,m\geq1$ for $A_u$ and $n\geq1,m\geq0$ for $A_Y$;
there is no $(0,0)$ mode. 

In contrast, applying the impedance boundary conditions in a PEC basis
couples the modal amplitudes through their fields at the walls. These
terms affect both attenuation and phase and are absent from independent
modal damping. Appendix~\ref{app:numerics} gives the coupled equations
and compares the attenuation coefficients.

\begin{figure*}[!tbp]
\includegraphics[width=\textwidth]{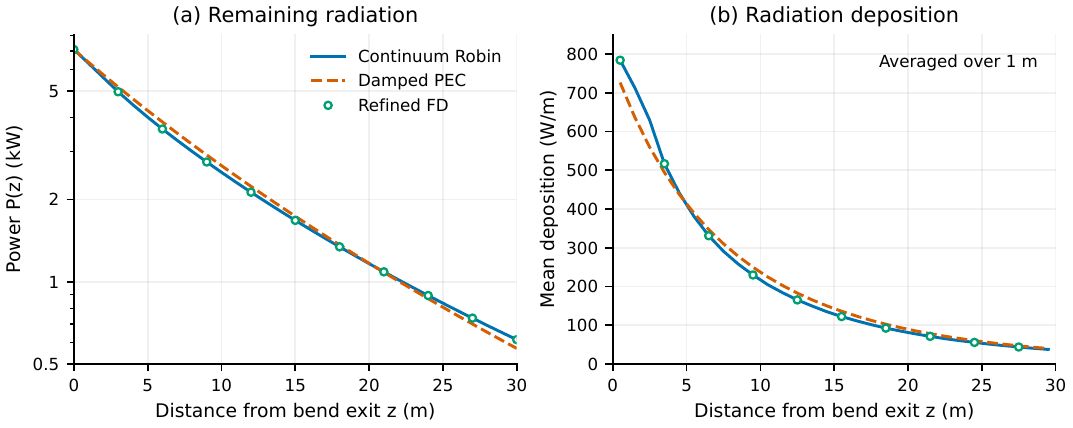}
\caption{Radiation transport in the 30 m conducting pipe without an outlet.
(a) Remaining power $P(z)$. (b) Radiation-only deposition $\bar q(z)$
[Eq.~\eqref{eq:intervaldeposition}], averaged over 1 m.
Distance $z$ is measured from the bend exit. All methods use the same
initial field; open markers show the FD results, nearly coincident
with Robin.}
\label{fig:absorption}
\end{figure*}

To show where the radiation is absorbed in Fig.~\ref{fig:absorption},
we calculate the mean power deposited per unit length over each 1 m
interval,
\begin{equation}
 \bar q(z)=\frac{P(z)-P(z+\Delta z)}{\Delta z},
 \qquad \Delta z=1\ {\rm m}.
 \label{eq:intervaldeposition}
\end{equation}
Thus $\bar q$ gives the average heat load over each 1 m interval,
without resolving every local interference maximum.

\begin{figure*}[!tbp]
	\includegraphics[width=\textwidth]{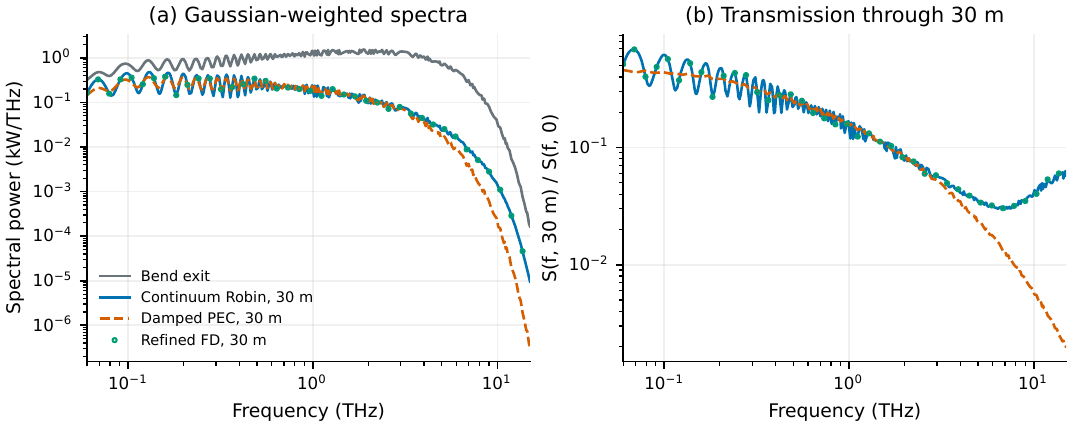}
	\caption{No-outlet spectrum. (a) Gaussian-weighted spectral power at the
		bend exit and after 30 m. (b) Transmission
		$S(f,30\,{\rm m})/S(f,0)$. Open circles show refined FD results.}
	\label{fig:baselinespectra}
\end{figure*}

For the bunch train in Table~\ref{tab:benchmark}, the no-outlet
calculation gives 7.09 kW at the bend exit in the 0.06--15 THz band.
The Robin solution deposits 6.48 kW over the next 30 m and leaves
0.61 kW in the pipe, compared with 0.57 kW for independently damped
PEC modes. The refined FD result agrees with the Robin solution
(Fig.~\ref{fig:absorption}). The difference from the damped-PEC result
is largest at high frequencies, where that approximation underestimates
the transmitted power (Fig.~\ref{fig:baselinespectra}). Refined FD and continuum Robin powers agree within
0.01\% at the sampled
frequencies. Refining the frequency grid changes a 1 m mean deposition
by at most 1.4\%; Appendix~\ref{app:numerics} gives these checks.

The oscillations in the bend-exit spectrum arise from excitation over
a finite bend length. Equation~\eqref{eq:mrresponse} contains
$4\sin^2[\lambda_\nu L_b/(4k)]/\lambda_\nu^2$ in each squared amplitude.
Frequency-dependent phase accumulation therefore produces oscillations
even without an outlet. This effect occurs within each mode and does not require interference
between different modes. Moving the observation plane through a
uniform PEC pipe cannot shift peaks of its integrated spectrum, although
the transverse field pattern changes. The spectrum in
Fig.~\ref{fig:baselinespectra} retains these oscillations. The oscillations persist under frequency and spatial refinement;
the frequency integral is checked separately
(Appendix~\ref{app:numerics}).

We also tested how the field differences in Sec.~\ref{sec:bendcontrol}
affect absorption. At 3 THz, replacing the paraxial bend-exit field by
the Helmholtz field changes the power deposited over 30 m by about
0.1\% and the transmitted power by less than 1\%. Both fields use the
same Robin propagation, with propagating modes from both scalar
families, odd $m\leq37$, and 1024 horizontal coefficients per family.
This test concerns absorption at one frequency, not broadband
extraction through the outlet.

Most of the radiation power is thus absorbed along the downstream
pipe. We now examine whether an outlet near the bend can divert part
of this radiation into a separate guide, treating the junction and
both outgoing guides as PEC.

\section{Extraction through a beam-free outlet}
\label{sec:fe}
\subsection{Geometry and moving finite elements}
The outlet has the same normal aperture as the beam pipe. Its straight
axis starts at the orbit point $s_0$ and makes angle $\theta_o$ with the
entrance direction. We measure its relative angle by
$\Delta\theta=\theta_o-\theta_{\rm tan}(s_0)$, where
$\theta_{\rm tan}(s_0)=s_0/R$ is the local orbit-tangent angle.
Both guide widths are measured normal to their own axes. The opening
and splitter are formed by the union of the two guides, with radius
$r=1.5$ mm at the upstream opening edge and the splitter nose.
The coordinate $s_0$ specifies the axis origin, not the nose position.
The outlet is treated as an outgoing radiation channel; the absorber
geometry and reflections from its termination are outside the model.

\begin{figure*}[!tbp]
\includegraphics[width=\textwidth]{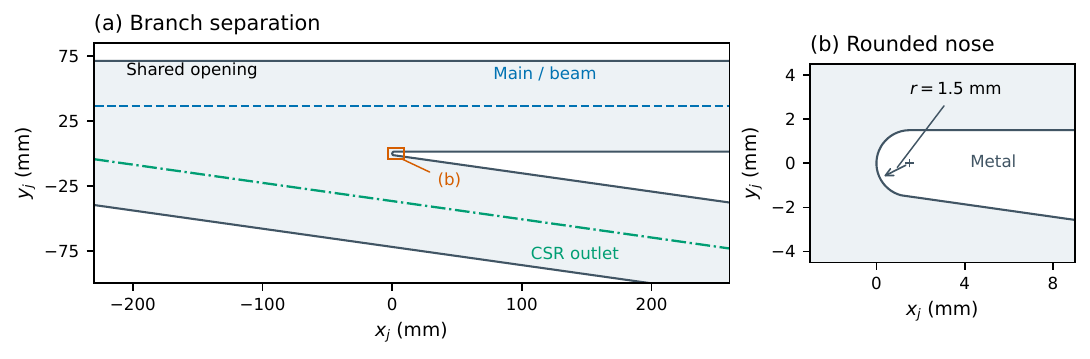}
\caption{Outlet junction: (a) separation of the beam and outlet guides;
(b) enlarged view of the splitter nose, with radius $r=1.5$ mm.
Coordinates are centered at the nose; $x_j$ follows the downstream beam
axis and $y_j$ points to its left.}
\label{fig:geometry}
\end{figure*}

The geometry illustrated in Figs.~\ref{fig:baselinegeometry}(a) and
\ref{fig:geometry} has $s_0=0.8L_b$ and an outlet directed $2^\circ$
below the local tangent, giving $\theta_o=22^\circ$.
The rounded nose lies about 0.38 m beyond the magnet. 

A section normal to the beam orbit contains one connected vacuum
interval before the splitter and two afterwards. We still write
$F=e^{\ii ks}(G+v)$, but choose a known field $G$ that does not depend
on the side-wall positions: the parallel-plate limit $d\to\infty$ of
Eq.~\eqref{eq:mrparticular},
\begin{align}
 G^E&=-C_m e^{-\alpha_m|u|},\nonumber\\
 G^H&=(C_m/\Z)\operatorname{sgn}(u)e^{-\alpha_m|u|}.
 \label{eq:knownfield}
\end{align}
The coefficient $C_m$ is unchanged. The parallel-plate field
$G$ satisfies the boundary conditions on the horizontal
plates but not on the side walls of the rectangular guide.
Consequently, unlike in Sec.~\ref{sec:fd}, the unknown $v$
contains both the radiation field and the side-wall correction
to the particular field $G$.

At the entrance to the bend ($s=0$), we set
$v(0,u)=G_{\mathrm{rect}}(u)-G(u)$,
where $G_{\mathrm{rect}}$ is the rectangular-guide particular
field defined in Eq.~\eqref{eq:mrparticular}.
Thus, the initial total field is
$F(0,u)=G_{\mathrm{rect}}(u)$,
identical to the equilibrium beam field used in the
no-outlet calculation, with no incident radiation.

Equation~\eqref{eq:metricmain} governs the remainder.
The electric PEC condition is $v^E=-G^E$ at each
side wall. For the magnetic scalar, the boundary condition
must account for the inclination of the walls relative
to the longitudinal coordinate $s$.
Writing the wall position as $u=u_w(s)$, the paraxial
boundary condition becomes
\begin{equation}
 \partial_u(G+v)=\frac{\ii k u_w'}{h_w^2}(G+v).
 \label{eq:movingwall}
\end{equation}
Here $u_w'=\dd u_w/\dd s$ and $h_w=1+\kappa u_w$.
It follows from the exact condition $\partial_n F^H=0$ on the
inclined wall after neglecting the longitudinal derivative of the
envelope. The full-wave comparisons in Sec.~\ref{sec:extraction}
check the effect of this approximation together with the paraxial
approximation in the interior.

To follow the changing side walls, we map each vacuum interval
onto a fixed interval $0\leq\xi\leq1$:
\begin{equation*}
 u=u_-(s)+w(s)\xi,
 \qquad w(s)=u_+(s)-u_-(s).
\end{equation*}
Here $u_-(s)$ and $u_+(s)$ are the wall positions,
$w(s)$ is the interval width, and $\xi$ is dimensionless.

We expand $v$ in continuous quadratic basis functions
$\phi_i(\xi)$. For the electric family, the boundary
condition $v^E=-G^E$ is imposed by introducing the
known function
\begin{equation*}
 B^E=-(1-\xi)G^E(u_-)-\xi G^E(u_+).
\end{equation*}
Then $v^E-B^E$ vanishes at both walls.
This boundary correction is called a lift.
For the magnetic family, we use $B^H=0$.

Let $\bm v$ and $\bm B$ contain the nodal values of the
remainder and the lift. With the scaled unknown
$\bm y=\sqrt{w}(\bm v-\bm B)$, the finite-element
discretization gives
\begin{equation*}
 D(s)\frac{\dd\bm y}{\dd s}
 =O(s)\bm y(s)+\bm f(s).
\end{equation*}
This is the system we integrate along the orbit.
Here $D$ is the mass matrix, $O$ is the evolution matrix,
and $\bm f$ is the source vector.

To specify the matrices, we define the weighted basis integrals
\begin{equation*}
\begin{aligned}
 M(g)_{ij}&=\int_0^1g\phi_i\phi_j\,\dd\xi,\\
 K(g)_{ij}&=\int_0^1g\phi_i'\phi_j'\,\dd\xi,\\
 C(g)_{ij}&=\int_0^1g\phi_i\phi_j'\,\dd\xi.
\end{aligned}
\end{equation*}
Here $g$ is a weight function, $i,j$ label FE nodes,
and primes on $\phi_i$ denote derivatives with respect
to $\xi$. The geometric factor and the change of the mesh
position along $s$ are
\begin{equation*}
\begin{aligned}
 h(s,\xi)&=1+\kappa\bigl[u_-(s)+w(s)\xi\bigr],\\
 v_w(s,\xi)&=
 \left.\frac{\partial u}{\partial s}\right|_\xi
 =u_-'(s)+w'(s)\xi.
\end{aligned}
\end{equation*}
Primes on $u_-$ and $w$ denote derivatives with respect
to $s$. The matrices in the evolution equation are
\begin{align}
 D={}&2\ii kM(h^{-1}),\nonumber\\
 O={}&\frac{K(h)}{w^2}+(\alpha_m^2-k^2)M(h)
       +k^2M(h^{-1})\nonumber\\
    &+\frac{2\ii k}{w}C(v_w/h)
       +\frac{\ii kw'}{w}M(h^{-1})+O_\Gamma.
 \label{eq:metricale}
\end{align}
The terms containing $v_w$ and $w'$ arise from the changing
transverse mesh. The boundary matrix $O_\Gamma$ accounts
for the inclination of the walls, while $\bm f$ contains
the residual source $r_G$ and contributions from the known
beam field and the electric boundary lift.
On smooth segments, the corresponding homogeneous
differential system conserves the weighted norm
$\bm y^\dagger M(h^{-1})\bm y$.

We integrate along $s$ using fourth-order methods.
In the bend, we use the two-stage implicit
Gauss--Legendre method. The matrices and source are
evaluated at two internal points of each step, including
the variation of the mass matrix $D(s)$.
In straight segments, $\kappa=0$ and
$D=2\ii kM(1)$ is constant in the mapped coordinates,
even when the transverse mesh changes.
There we use a commutator-free fourth-order exponential
integrator (CF4)~\cite{BlanesMoan2006}.
Both methods advance the same paraxial system.
Steps end at changes of curvature, across which we impose
continuity of the envelope. Reflected waves and exact
interface matching are included only in the separate
Helmholtz reference solution.

The splitter tip requires special treatment because a
transverse section changes from one vacuum interval to two.
The rounded nose lies in the straight continuation.
Near its tip, the two intersections of a transverse
section with the nose contour satisfy
\begin{equation*}
 u_\pm=u_*\pm\sqrt{2r(s-s_*)}+O(s-s_*),
\end{equation*}
where $(s_*,u_*)$ is the tip position and $r$ is the
nose radius.
Although the contour is smooth, $du_\pm/ds$ diverges
as $s\to s_*^+$. A small-wall-slope assumption therefore
does not hold arbitrarily close to the tip.
We place a step boundary at $s=s_*$ and evaluate
subsequent quadrature points strictly downstream,
avoiding these singular derivatives at the tip.

At this point, we transfer $v$ from the original mesh
to the two new meshes by an $L^2$ projection.
The projection minimizes the integrated squared
difference between the old and new field representations.
The electric boundary values are imposed during the
projection; the magnetic boundary condition enters
the subsequent evolution.
Appendix~\ref{app:numerics} gives the explicit boundary
and source terms, the conservation identity, the integration
and projection formulas, and the convergence checks.

\subsection{Radiation partition and extraction}
\label{sec:extraction}
We evaluate the branch powers after the two guides have separated and
become uniform. For the illustrated geometry, the sampling sections
are 10 mm beyond the end of the rounded nose and normal to the beam
orbit. We use the paraxial norm in Eq.~\eqref{eq:paraxialpower} on
each branch of these sections. Further propagation in the uniform PEC
guides changes modal phases but not the integrated power in either
branch.

\begin{figure*}[!tbp]
\includegraphics[width=\textwidth]{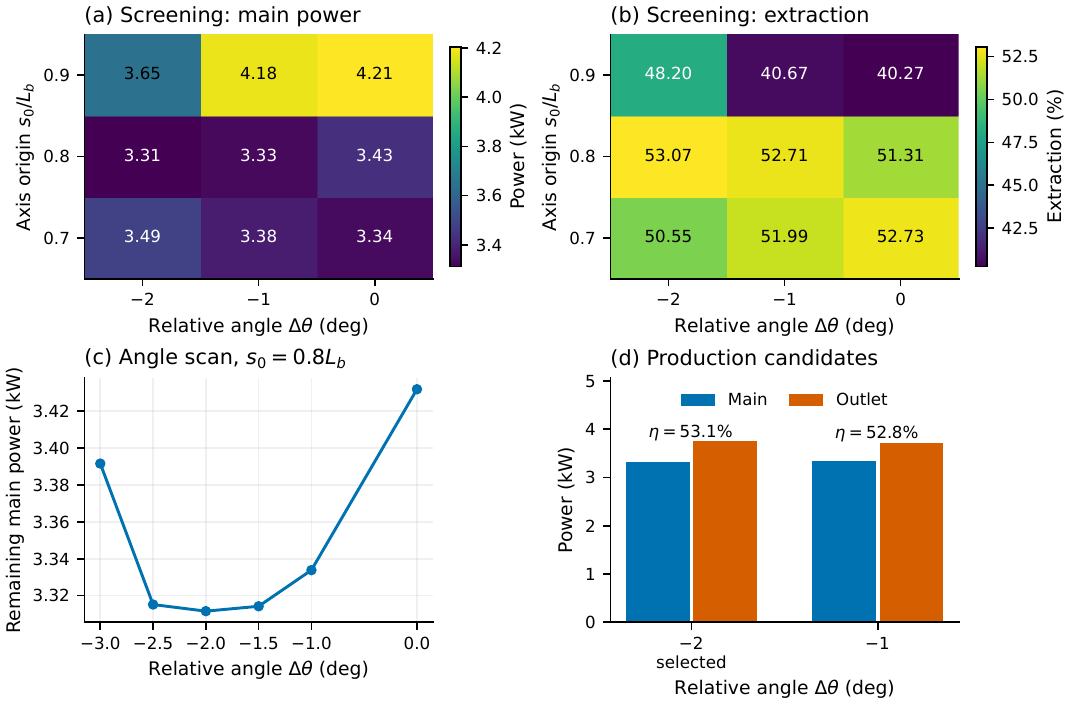}
\caption{Dependence on outlet geometry: (a,c) main-guide power $P_b$;
(b) extraction fraction $\eta$; (d) powers in both guides.
Panels (a,b) show the preliminary scan; (d) compares the two final
candidates. The relative angle is $\Delta\theta=\theta_o-\theta_{\rm tan}(s_0)$.
In (c,d), $s_0=0.8L_b$, so $\Delta\theta=-2^\circ$ gives
$\theta_o=22^\circ$. The selected geometry minimizes $P_b$ within
the family studied.}
\label{fig:survey}
\end{figure*}

For power evaluation, the radiation envelope in the beam channel is
$G+v-G_{\rm rect}$, where $G_{\rm rect}$ is the rectangular particular
field in Eq.~\eqref{eq:mrparticular}. In the beam-free outlet it is
$G+v$. Substituting these envelopes for $v$ in
Eqs.~\eqref{eq:paraxialpower} and \eqref{eq:bunchpower} gives the average
powers $P_b$ and $P_o$ in the beam and outlet branches, respectively.
The subscript $b$ here denotes the beam branch. We vary the outlet
origin and angle at fixed apertures
and rounding radii to minimize $P_b$. We also report the extraction
fraction
\begin{equation}
 \eta=\frac{P_o}{P_b+P_o}.
 \label{eq:eta}
\end{equation}
The denominator is the two-branch power of that geometry, not the power
of a different no-outlet problem. The opening changes the beam-driven
field, so the two totals need not coincide even with PEC walls.
The survey first samples three origins, $s_0/L_b=0.7,0.8,0.9$, and
three relative angles, $\Delta\theta=-2^\circ,-1^\circ,0$. An extended
scan at $s_0=0.8L_b$ reaches $-3^\circ$ and reveals a shallow minimum of
main-channel power near $-2^\circ$ (Fig.~\ref{fig:survey}). The preliminary scan and final comparison use different resolutions
and are shown separately. Both final candidates use the same spatial
and frequency resolution.
The selected angle lowers main power by about 22 W relative to
$-1^\circ$, compared with a 2 W change in this difference under frequency
refinement. The difference is resolved by this refinement test, but the shallow
minimum does not establish a unique or global optimum.

\begin{figure*}[!tbp]
\includegraphics[width=\textwidth]{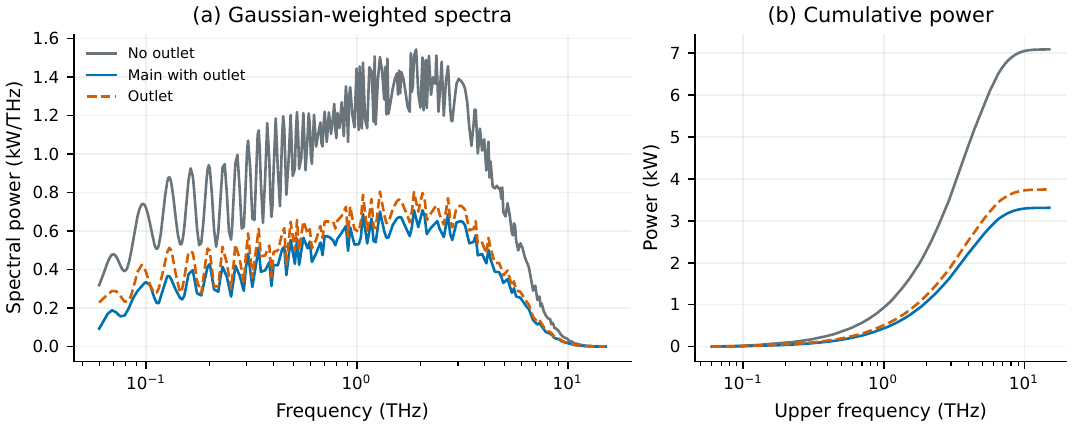}
\caption{Radiation from the no-outlet pipe and the two guides of the selected
outlet. (a) Gaussian-weighted spectra at the bend exit without an outlet
and after guide separation with an outlet. (b) Cumulative power
$\int_{f_{\min}}^f S(f')\dd f'$, with $f_{\min}=0.06$ THz.}
\label{fig:outletspectra}

\end{figure*}

\begin{table}[tb]
\caption{Broadband radiation partition for 60 pC Gaussian bunches
at 9.75 mA. Powers $P_b$ and $P_o$, in kW, are carried by the main and
outlet branches in the PEC geometry. The no-outlet value is the power
at the bend exit. Outlet cases use $s_0=0.8L_b$; angles are relative
to the local tangent. The extraction fraction $\eta=P_o/(P_b+P_o)$
uses each outlet geometry's own two-branch power.}
\label{tab:outletpowers}
\begin{ruledtabular}\begin{tabular}{lrrr}
Geometry & $P_b$ & $P_o$ & $\eta$ (\%) \\
No outlet & 7.09 & --- & --- \\
Outlet, $\Delta\theta=-1^\circ$ & 3.33 & 3.72 & 52.8 \\
Outlet, $\Delta\theta=-2^\circ$ & 3.31 & 3.75 & 53.1 \\
\end{tabular}\end{ruledtabular}
\end{table}

For the selected geometry, 3.75~kW of radiation power enters the
beam-free outlet, while 3.31~kW remains in the beam channel
(Table~\ref{tab:outletpowers}). The resulting extraction fraction is 53.1\%.  The outlet and main powers correspond to approximately 385 and
340 W per milliampere, respectively, when the repetition rate is varied
at a fixed bunch charge of 60 pC. The spectra in
Fig.~\ref{fig:outletspectra} show how that division varies across the
Gaussian-weighted band. The no-outlet curve is a separate beam-driven
solution; it is not obtained by adding the two outlet branches.

Halving the element and step lengths at selected frequencies changes
the branch powers by less than 0.1\%; refining the frequency grid
changes the integrated powers by less than 1\%. The complex fields
are more sensitive to resolution, so this power convergence should
not be interpreted as convergence of every field component.
Appendix~\ref{app:numerics} gives the resolutions and field comparisons.
These numerical tests do not assess the paraxial approximation itself.

\begin{figure*}[!tbp]
\includegraphics[width=\textwidth]{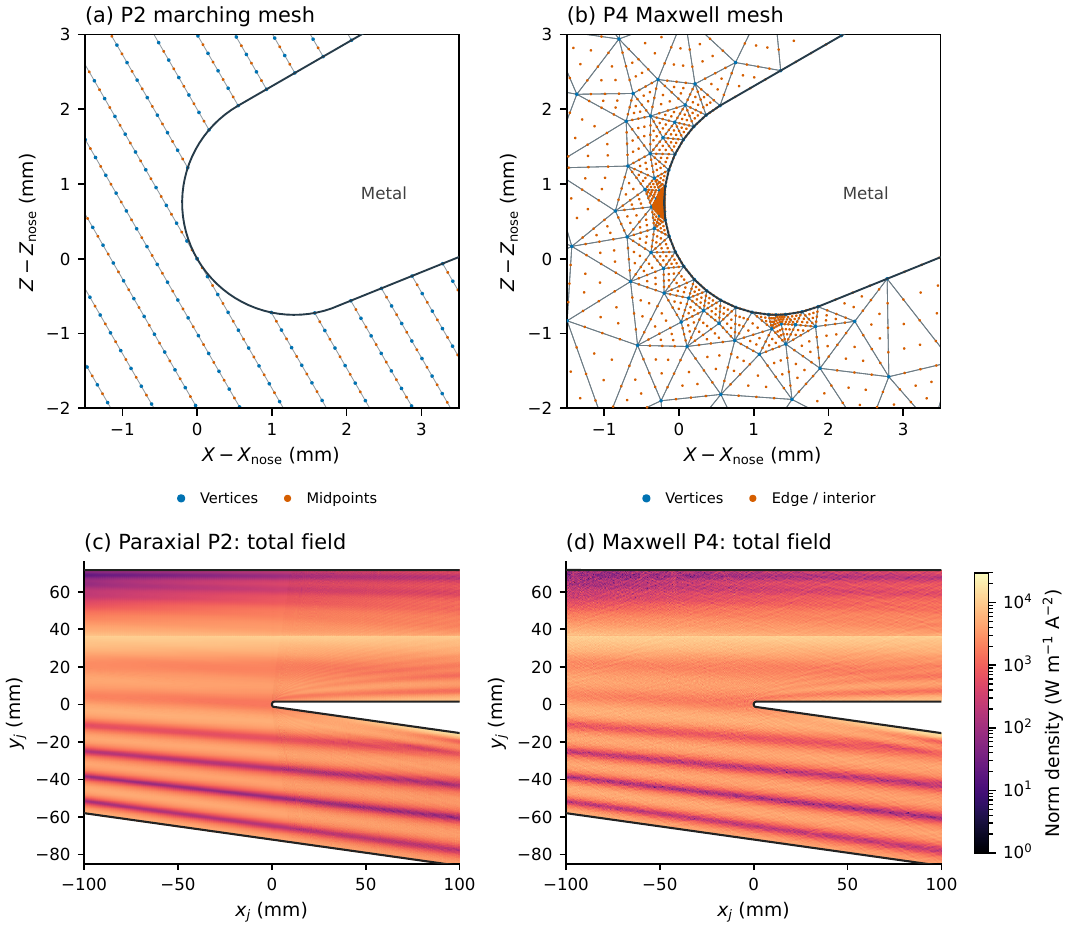}
\caption{Meshes and total fields near the splitter. (a) Paraxial P2 mesh at
0.241 THz. (b) Full-wave P4 mesh at 0.24 THz. Blue points are vertices;
orange points are additional element nodes. (c,d) P2 and P4 fields at
0.24 THz, $m=1$, shown as $b(|F^E|^2/\Z+\Z|F^H|^2)/(4|I_\omega|^2)$
on common scales. This quantity includes the beam field and is not the
local Poynting flux. P2 uses a line source; P4 uses Gaussian widths
$\sigma_Z=25\,\mu$m and $\sigma_Y=80\,\mu$m.}
\label{fig:meshfields}
\end{figure*}

The modal matching used for the constant-cross-section bend in
Sec.~\ref{sec:bendcontrol} is not readily applicable to the branching
geometry. We therefore check the paraxial calculation with a full-wave
solution using quartic (P4) triangular finite elements.
Figure~\ref{fig:meshfields} shows the meshes and total fields for the
two methods.
As in the modal approach, vertical separation gives
two scalar Helmholtz problems, now solved on the full
horizontal domain.
The walls are PEC, and the ports permit outgoing
radiation and evanescent decay.
We include the known phase $e^{\ii ks}$ in the FE basis to reduce the
oscillation that the mesh must resolve. No longitudinal derivatives
are dropped.
Unlike the paraxial calculation, which uses a transverse
line source with analytical subtraction of the singular
beam field, the P4 calculation uses a conserved current
with finite Gaussian transverse widths $\sigma_Z$ in
global-$Z$ slices and $\sigma_Y$ vertically.
We therefore test mesh and source-width sensitivity
separately.

For this full-wave comparison, we use common planes normal to the
individual outgoing guide axes, rather than the orbit-normal sections
used for the broadband paraxial calculation. We restore the carrier phase, subtract the beam field in
the main guide, and project onto that port's sine and cosine functions.
The outlet contains no beam field to subtract. Equation~\eqref{eq:portflux}
then gives the outgoing flux using the normal guide width. For the paraxial field, this assigns the exact outgoing-mode weights
to its projected amplitudes; it does not make the field and its normal
derivative satisfy the full Maxwell equations. The broadband results above use the paraxial norm
on orbit-normal sections instead. Appendix~\ref{app:maxwell} supplies
the Maxwell source, assembly, and projection details.

At 0.24 THz ($m=1$), the paraxial P2 and full-wave P4 calculations
differ in radiation power by about 0.3\% in the main guide and 1.2\%
in the outlet. Changes under P4 mesh refinement and reduction of the
Gaussian source width are smaller than the outlet difference.
The agreement supports the predicted power division at this frequency,
but the different sources prevent us from attributing the entire
difference to the paraxial approximation.

A second comparison tests scattering at the splitter separately from
radiation excitation in the bend. Near 3.4 THz ($m=1$), we restrict
the domain to a short region around the nose and attach a uniform
input guide 10 mm upstream of it. A separate paraxial calculation
of the upstream bend and opening supplies identical complex
propagating input modes to the P2 and P4 calculations.
The branch powers agree closely, and Maxwell predicts percent-level
reflection. However, the artificial straight inlet creates a corner
where it meets the inclined wall, and spatial convergence of this
local Maxwell calculation has not been established. This is therefore a check of scattering for the prescribed incident
field in the modified geometry, not an error estimate for the complete
beam-driven problem.

The PEC calculation quantifies radiation available to an absorber
in the outlet. It does not calculate heating in the junction or in the
continuing main pipe. Predicting that heating would require the complex
field after separation and a specified finite-conductivity geometry;
the extraction fraction alone is insufficient.

\section{Conclusions}
\label{sec:conclusions}

We have proposed a beam-free side outlet to divert CSR from a bent
beam pipe. To calculate the generated radiation and its division between
the two guides, we extended the Stupakov--Kotelnikov modal solution to
Eq.~\eqref{eq:metricmain} and developed FD and quadratic FE solutions
of the same equation. The FE method follows the changing cross section
through the branching region.

The additional curvature terms improve agreement with the full-wave
bend solution. The three numerical methods agree closely, but comparison
with full-wave solutions still shows phase and profile differences,
especially at high frequencies. Good agreement in total power therefore
does not establish equal accuracy of the fields, and the full-wave
checks at selected frequencies do not give an error bound over the
entire spectrum.

For the conducting pipe without an outlet, we developed a complex mode
expansion satisfying impedance boundary conditions. It includes changes
in the mode profiles and propagation constants, as well as interference
between modes in the power. Agreement with refined FD calculations
checks the expansion within the chosen wall model. For this example,
independently damped PEC modes underestimate the transmitted power at
high frequencies.

For the selected PEC geometry and bunch parameters, the outlet carries
3.75~kW, or about 53\% of the total outgoing radiation power, while
3.31~kW remains in the beam pipe. The dependence on outlet position
and angle has a shallow minimum in the remaining beam-channel power
within the family studied. Thus a substantial fraction of CSR can be
directed into a beam-free channel. 

\begin{acknowledgments}
OpenAI Codex assisted with literature checking, mathematical exposition, and
  manuscript preparation.
\end{acknowledgments}
\appendix
\section{Numerical details}
\label{app:numerics}
This appendix gives the FD discretization for the conducting pipe,
the paraxial reduction, the FE implementation, and convergence checks.

\subsection{Straight-guide propagation}
\label{sec:wallnumerics}
For the FD solution of Eq.~\eqref{eq:robin}, the normal electric-field
component is stored at cell centers, half a cell from the wall. With
spacing $\Delta x$ and $g=\ii k\chi\Delta x/2$, the ghost value and
wall value are
\begin{equation}
 A_g=\frac{1+g}{1-g}A_i,
 \qquad A_w=\frac{A_i}{1-g}.
 \label{eq:fdghost}
\end{equation}
Here $A_i$, $A_g$, and $A_w$ are the adjacent interior, ghost, and wall
values. The transverse second-difference matrix has interior diagonal
$-2/\Delta x^2$, adjacent entries $1/\Delta x^2$, and endpoint diagonal
$[-2+(1+g)/(1-g)]/\Delta x^2$. Tangential components have Dirichlet endpoints.
We diagonalize this complex symmetric matrix independently of the
continuum Robin eigenproblem. The discrete power balance uses $A_w$,
so the wall loss contains the factor $|1-g|^{-2}$.

Dense diagonalization can be avoided without changing the FD operator.
For $N=D/\Delta x$ cells, the even and odd eigenvectors satisfy
\begin{align}
 \sin(qD/2)\sin(q\Delta x/2)
 +g\cos(qD/2)\cos(q\Delta x/2)&=0,\nonumber\\
 \cos(qD/2)\sin(q\Delta x/2)
 -g\sin(qD/2)\cos(q\Delta x/2)&=0.
 \label{eq:discreterobinroots}
\end{align}
respectively. Their eigenvalues are $-4\sin^2(q\Delta x/2)/\Delta x^2$.
Normalization, projection, and Gram matrices use the discrete inner
product $\Delta x\sum_j$. For the cell centers
$x_j=-D/2+(j+1/2)\Delta x$, $j=0,\ldots,N-1$, and any wave number $q$,
the exact sum
\begin{equation}
 \Delta x\sum_j e^{\ii q x_j}
 =D\frac{\operatorname{sinc}(qD/2)}{\operatorname{sinc}(q\Delta x/2)},
 \quad \operatorname{sinc}x=\sin x/x,
\end{equation}
evaluates the required overlaps. We check this calculation against
direct matrix exponentials on small grids and increase both grid size
and retained mode count for the broadband results.

For the source-free bend FD system in Sec.~\ref{sec:bendmethod},
a mode associated with an eigenvalue $\lambda$ of $B$ propagates by
$\exp[\ii\lambda\Delta s/(2k)]$. The CN multiplier is
\begin{equation}
 r_{\rm CN}=\frac{1+\ii\Delta s\lambda/(4k)}
                  {1-\ii\Delta s\lambda/(4k)}.
 \label{eq:mrcn}
\end{equation}
Its modulus is also unity, but its phase differs from the exact
exponential. Comparing the two on the same grid therefore isolates
the longitudinal stepping error.

For the conducting pipe, let $\lambda$ instead denote an eigenvalue
of the FD evolution operator. Exact
propagation uses $e^{\lambda z}$. For a numerical step $\delta z$,
CN has multiplier $r_{\rm CN}=(1+\delta z\lambda/2)/(1-\delta z\lambda/2)$.
The step $\delta z$ is distinct from the averaging length $\Delta z$ in
Eq.~\eqref{eq:intervaldeposition}. If $|\delta z\lambda|\ll1$, then
\begin{equation}
 \frac{\log r_{\rm CN}}{\delta z}
 =\lambda+\frac{\delta z^2\lambda^3}{12}+O(\delta z^4\lambda^5).
 \label{eq:cnmodified}
\end{equation}
CN stability does not guarantee accurate phase or damping over 30 m.
For poorly resolved modes with a large imaginary eigenvalue, its
multiplier has modulus close to one even when the exact mode decays.
We therefore use exact propagation of the FD system for the broadband
results.

The distinction between power and field convergence is especially clear
for CN propagation. Near 3 THz, a $50\,\mu$m step gives less than
0.04\% power error after 30 m, but about 81\% complex-field error
relative to exact evolution on the same FD grid. Reducing the step
eightfold lowers the field difference to about 2\%. The large field error is therefore mainly an accumulated phase error.

For comparison, let $c_{nm}^j$ be the coefficients of component $j$ in
an orthonormal PEC basis. With the independent attenuation defined in
Sec.~\ref{sec:absorption}, orthonormality gives
\begin{equation}
 p_{\rm PEC}=\frac1{2\Z}\sum_{j,n,m}|c_{nm}^j(0)|^2
                                      e^{-2\alpha_{nm}^{\rm PEC}z}.
\end{equation}

To explain the difference between independent PEC-mode
damping and the impedance model, consider the expansion
of an electric-field component in orthonormal PEC modes,
$A_j=\sum_\nu c_\nu(z)\Phi_\nu$.
Here $\nu=(n,m)$ is the combined mode index, and
$k_{c,\nu}^2=(n\pi/a)^2+(m\pi/b)^2$.
Projecting the propagation equation with the impedance boundary
conditions onto this basis gives
\begin{equation}
c_\nu'=-\frac{\ii k_{c,\nu}^2}{2k}c_\nu
-\frac{\chi}{2}\sum_\mu B_{\nu\mu}c_\mu,
\qquad
B_{\nu\mu}=\int_{\Gamma_j}\Phi_\nu^*\Phi_\mu\,\dd\ell.
\label{eq:pec_mode_coupling_app}
\end{equation}
Here the prime denotes differentiation with respect to
$z$, and $B_{\nu\mu}$ describes the overlap of PEC modes
at the walls. Modes orthogonal over the cross-sectional
area need not be orthogonal on the boundary.
The off-diagonal terms therefore couple their amplitudes.
Moreover, the prescribed attenuation coefficients in
Eq.~\eqref{eq:prescribedattenuation} are not obtained simply by retaining the
diagonal terms of this matrix.

For the electric-field component $A_u$, the wall-overlap
matrix has the explicit form
\begin{equation}
 B^{(u)}_{nm,n'm'}=
 \frac{\sqrt{(2-\delta_{n0})(2-\delta_{n'0})}}a
 [1+(-1)^{n+n'}]\delta_{mm'}.
\end{equation}
Keeping only its diagonal and $\operatorname{Re}\chi$ gives amplitude loss
$\gamma^u_{nm}=(2-\delta_{n0})\operatorname{Re}\chi/a$ and
$\gamma^Y_{nm}=(2-\delta_{m0})\operatorname{Re}\chi/b$, which differ from the
prescribed
$\alpha_{nm}^{\rm PEC}$. Thus independent PEC damping differs from the impedance model both in
the attenuation coefficients and in the omitted coupling and phase shifts.
Adding more PEC modes does not remove this difference.

For the common no-outlet field at 3.398 THz, the fraction surviving
30 m is about 5.0\% with either Robin or refined FD, compared with
4.2\% for diagonal Robin loss and 4.3\% for the prescribed PEC loss.
The diagonal Robin calculation separates the effect of neglecting mode
coupling from that of changing the attenuation coefficients.

These comparisons test solutions of the same impedance model. The
full-wave calculations use PEC walls and do not test the
finite-conductivity boundary approximation.

At the sampled frequencies, refined FD and Robin powers after 30 m
agree within 0.01\%; basis refinement, projection, and power-balance
checks give consistent results. Frequency integration is more sensitive:
increasing the no-outlet grid from 160 to 319 frequencies changes the
mean deposition in a 1 m interval by at most 1.4\%.

\subsection{Paraxial reduction and FE propagation}
\label{sec:fenumerics}
The scalar orbit Laplacian is
$h^{-1}\partial_s(h^{-1}\partial_s)+h^{-1}\partial_u(h\partial_u)$.
Let $e^{\ii ks}S_m$ denote the source in the separated scalar Helmholtz
equation $(\Delta+k^2-\alpha_m^2)F=e^{\ii ks}S_m$, where $\Delta$ is this
horizontal Laplacian. For constant curvature, the envelope
$\widetilde F=e^{-\ii ks}F=G+v$ satisfies
\begin{equation}
 \begin{split}
 \widetilde F_{ss}+2\ii k\widetilde F_s
       +h^2(\widetilde F_{uu}-\alpha_m^2\widetilde F)&\\
       +\kappa h\widetilde F_u+k^2(h^2-1)\widetilde F&=h^2S_m.
 \end{split}
 \label{eq:exactmetric}
\end{equation}
Subscripts denote partial derivatives. Neglecting $\widetilde F_{ss}$
and subtracting $G$ gives Eq.~\eqref{eq:metricmain} within each
constant-curvature segment. The treatment of interfaces is described
in Sec.~\ref{sec:fe}.
The beam-field subtraction removes the singular source terms
analytically. The magnetic particular field has derivative
\begin{equation*}
 \frac{\dd G^H}{\dd u}=(G^H)'_{\rm reg}
                  +\frac{2C_m}{\Z}\delta(u),
\end{equation*}
where $\delta$ is the Dirac distribution and the regular part is the
ordinary derivative away from $u=0$. The first derivative of $G^E$
contains no delta term, but its second derivative does.
For the line current, with $\delta'$ denoting the derivative of the
Dirac distribution, the magnetic curl source is
$(2C_m/\Z)[\delta'(u)+\kappa\delta(u)]$ before multiplication by $h^2$.
Since $u\delta'=-\delta$, it becomes
$(2C_m/\Z)[\delta'(u)-\kappa\delta(u)]$ afterwards.
Subtracting the known field in Eq.~\eqref{eq:knownfield} cancels these
delta-function terms and leaves $r_G$ in Eq.~\eqref{eq:metricmain}.

For comparison, let $q_\perp$ be a characteristic transverse wave
number and $\epsilon\ll1$ an ordering parameter. The leading paraxial
ordering takes
$q_\perp/k=O(\epsilon^{1/2})$,
$\widetilde F_s/(k\widetilde F)=O(\epsilon)$, and
$\kappa u=O(\epsilon)$. Diffraction and $2k^2\kappa u\widetilde F$
then appear at leading order; $\widetilde F_{ss}$,
$2\kappa u(\widetilde F_{uu}-\alpha_m^2\widetilde F)$,
$\kappa\widetilde F_u$, and $k^2\kappa^2u^2\widetilde F$ are smaller
in the local ordering.
Equation~\eqref{eq:metricmain} retains these geometric terms but drops
the longitudinal second derivative. It is therefore not a complete
higher-order expansion; the full-wave comparisons test the resulting
phase and boundary errors.

The exact oblique magnetic wall condition is
\begin{equation}
 \widetilde F_{u}-\frac{u_w'}{h_w^2}(\ii k\widetilde F+\widetilde F_{s})=0.
\end{equation}
Dropping $\widetilde F_s$ gives Eq.~\eqref{eq:movingwall}.

For the P2 discretization, the local coordinate
$0\leq\tau\leq1$ gives the three quadratic shape
functions $2\tau^2-3\tau+1$, $4\tau(1-\tau)$,
and $2\tau^2-\tau$.

The boundary matrix $O_\Gamma$ vanishes for the
electric family. For the magnetic family, its
nonzero boundary entries are
\begin{equation*}
\begin{aligned}
 (O_\Gamma)_{00} &= \frac{\ii k u'_-}{w h_-},\\
 (O_\Gamma)_{NN} &=-\frac{\ii k u'_+}{w h_+},
\end{aligned}
\end{equation*}
where $0$ and $N$ label the endpoint nodes and
$h_\pm=1+\kappa u_\pm$.

The volume contribution to the source vector is
\begin{equation*}
 f_i^{\mathrm{vol}}=\sqrt{w}\int_0^1\phi_i r_G\,\dd\xi.
\end{equation*}
For the magnetic family, the additional endpoint
contributions are
\begin{equation*}
 \frac{-h_-G'(u_-)+\ii k u'_-G(u_-)/h_-}{\sqrt{w}}
\end{equation*}
at the left boundary and
\begin{equation*}
 -\frac{-h_+G'(u_+)+\ii k u'_+G(u_+)/h_+}{\sqrt{w}}
\end{equation*}
at the right boundary.
For the electric family, the boundary lift contributes
\begin{equation*}
 \sqrt{w}\left\{O\bm B-D\left[
 \bm B'+\frac{w'}{2w}\bm B\right]\right\}.
\end{equation*}
The term proportional to $w'$ is required for
equivalence with the unscaled moving-interval equation.

Away from topology and curvature changes, the
source-free system conserves
$\bm y^\dagger M(h^{-1})\bm y$.
In the scaled coordinates, the corresponding
matrix identity is
\begin{equation}
 \frac{\dd M(h^{-1})}{\dd s}
 +\frac{O-O^\dagger}{2\ii k}=0.
 \label{eq:fe_norm_balance_app}
\end{equation}
The identity includes the changing boundaries and mass matrix and
holds to roundoff in separate straight- and curved-interval tests.
It does not apply when the new PEC boundary is introduced at the splitter.

Eight-point element quadrature evaluates the matrices and loads in
Eq.~\eqref{eq:metricale}. We advance the curved segment with a two-stage Gauss method,
evaluating the mass matrix separately at each stage. Its stage
locations are $c_1=1/2-\sqrt3/6$ and
$c_2=1/2+\sqrt3/6$, and its coefficients are
$a_{11}=a_{22}=1/4$,
$a_{12}=1/4-\sqrt3/6$, and $a_{21}=1/4+\sqrt3/6$. For a step from $s_n$
to $s_n+\Delta s$, subscripts $i=1,2$ on $D$, $O$, and $\bm f$ denote
evaluation at $s_n+c_i\Delta s$. The stage slopes $\bm K_i$ obey
\begin{align}
 D_i\bm K_i-\Delta s\sum_{j=1}^2a_{ij}O_i\bm K_j
     &=O_i\bm y_n+\bm f_i,\nonumber\\
 \bm y_{n+1}&=\bm y_n+\frac{\Delta s}{2}(\bm K_1+\bm K_2).
 \label{eq:metricgauss}
\end{align}
Interleaving the two stage vectors gives a banded system with half
bandwidth five. After the boundary lift, the electric endpoint
coefficients are zero, as are their stage derivatives. The method is
fourth order on smooth segments.

In straight segments $D=2\ii kM$ is constant on the reference interval,
where $M=M(1)$ is the unweighted mass matrix.
To include the source in the CF4 step, we use the augmented vector
$\widehat{\bm y}=(\bm y,1)^T$. Let $L_1$ and $L_2$ be the matrices
that represent $\bm y'=D^{-1}(O\bm y+\bm f)$ in these augmented
coordinates at the two Gauss locations. Then
\begin{align}
 a_1&=(3-2\sqrt3)/12,\qquad a_2=(3+2\sqrt3)/12,\nonumber\\
 \widehat{\bm y}_{n+1}&=e^{\Delta s(a_1L_1+a_2L_2)}
              e^{\Delta s(a_2L_1+a_1L_2)}\widehat{\bm y}_n.
\end{align}
The rightmost exponential acts first. For a scalar argument $z$, the
$[3/3]$ Pad\'e approximation to $e^z$ is
\begin{equation}
 R_3(z)=\frac{120+60z+12z^2+z^3}{120-60z+12z^2-z^3}
\end{equation}
Each matrix exponential is then evaluated through three banded
linear solves. Local elimination
of P2 midpoints leaves a tridiagonal vertex system, or a tridiagonal
system of two-by-two stage blocks for coupled Gauss propagation.
Constant-section matrices are factored once and reused.

At first contact with the splitter, an $L^2$
projection transfers the remainder field from
the parent interval to the two child intervals.
Let $v_p$ denote the parent field and $\phi_i^c$
the basis functions on a child interval.
The projection load and child mass matrix are
\begin{equation*}
\begin{aligned}
 (b_c)_i &=\int \phi_i^c v_p\,\dd u,\\
 (M_c)_{ij} &=\int \phi_i^c\phi_j^c\,\dd u,
\end{aligned}
\end{equation*}
where the integrals extend over the child interval.

For the electric family, the prescribed boundary
values are imposed during the projection.
With $I$ and $B$ denoting the interior and boundary
degrees of freedom, respectively, the coefficients satisfy
\begin{equation*}
\begin{aligned}
 M_{II}\bm v_I &=\bm b_I-M_{IB}\bm g_B,\\
 \bm v_B &=\bm g_B=-\bm G_B.
\end{aligned}
\end{equation*}
Here $M_{II}$ and $M_{IB}$ are blocks of the child mass
matrix, $\bm b_I$ is the interior part of the projection
load, and $\bm G_B$ contains the known beam field
at the new endpoints.

For the magnetic family, the coefficients satisfy
\begin{equation*}
 M_c\bm v_c=\bm b_c.
\end{equation*}
The magnetic derivative boundary condition enters
the weak evolution equation rather than the projection.
Integrating over intersections of parent and child
elements gives exact products of the P2 basis functions.
The electric boundary condition can change the weighted norm.
We record this change without renormalization and do not assume
fourth-order accuracy at the split.

We refine the spatial and frequency grids separately. Using the
transverse formation scale $\ell_f=(R/k^2)^{1/3}$, we choose a P2
element length no greater than
$\min(1\,{\rm mm},\ell_f/16)$ through 6 THz and $\ell_f/12$ above it.
Below that transition the longitudinal step satisfies
\begin{equation}
 \Delta s\leq\min\left(0.125\,{\rm mm},\frac{0.25R}{ka}\right);
\end{equation}
above 6 THz these step limits are doubled. Steps end at changes in
curvature, the opening, and the splitter tip. The highest odd vertical
index increases from 9 to 75 across the frequency band. The preliminary
geometry scan uses 12 frequencies, element length $\ell_f/12$, and the
doubled step limits. Both final candidates use the same resolution.

The no-outlet horizontal basis starts at 160 functions per family and
increases with frequency; the last three retained vertical modes carry
less than $3\times10^{-5}$ of the entrance norm. Independent bend tests
show second-order FD convergence. At 0.24 THz for the full retained
field, and at 3 THz for $m=1,3$, FD and P2 field differences from the modal solution remain below
0.05\% and 0.01\%, respectively. We use Eq.~\eqref{eq:benderrors}
with the modal solution of Eq.~\eqref{eq:metricmain} as the reference,
summing all retained vertical modes for the full field.

Broadband powers are integrated by the trapezoidal rule in physical
frequency over 0.06--15 THz, using 160 frequencies for the outlet and
319 for the no-outlet reference. Refinement from the respective coarser
grids changes these powers by less than 1\%. Dense scans near 0.14, 1,
and 3.4 THz, with fixed spatial resolution and separate refinement at
spectral extrema, preserve the oscillations. Extending the no-outlet
band down to 6.1 GHz and up to 20 THz adds less than 0.1\% to its
integral. This bandwidth test applies only to the no-outlet spectrum. It does
not estimate the omitted outlet power or the contribution above
20 THz. The paraxial approximation also becomes unreliable near the
lowest propagation cutoff.

Outlet refinement halves both element and step lengths at four
frequencies between 0.24 and 6.21 THz and increases the vertical cutoff
at the first three. Branch powers change by less than 0.1\%. All
refinement sensitivities in this appendix are measured relative to
the coarser result.
The complex fields change by as much as 43\%. The largest differences
come from transverse components with $q_n^2+\alpha_m^2\geq k^2$,
where $q_n$ is the horizontal wave number. The paraxial equation
propagates these components, whereas in the Helmholtz equation they
are evanescent or at cutoff. They contribute at most 2\% of the refined
norm. Using only propagating components and the flux norm reduces the
largest field difference to 14\%. The reported PEC calculations retain
the full field.

These tests at selected frequencies do not give a field-error bound
for the whole band. We also check port quadrature and transfer at the
tip separately. At 0.24 THz, moving the ports by 0.1 m changes outgoing
powers by less than 0.2\%.

\section{Independent Maxwell and Helmholtz formulations}
\label{app:maxwell}
We use the vertical separation in Eq.~\eqref{eq:verticalbasis} for
both full-wave calculations: modal matching in the rectangular bend
and finite elements in the branching geometry.

\subsection{Bend with a common line source}
We solve Eq.~\eqref{eq:helmholtzcontrol} with the interface and radiation
conditions stated in Sec.~\ref{sec:bendcontrol}. The field $U$ is already
the radiation after subtraction of $e^{\ii ks}G$.

Expand $U$ in normalized straight-guide sine (E) or cosine (H) functions
$\phi_n$, including the zero horizontal magnetic index. With $d=a/2$
and primes on $\phi_n$ denoting $u$ derivatives, define
\begin{align}
 W_{ij}&=\int_{-d}^d\phi_i\phi_j/h\dd u,\nonumber\\
 A_{ij}&=\int_{-d}^dh[(k^2-\alpha_m^2)\phi_i\phi_j
                                  -\phi_i'\phi_j']\dd u,\nonumber\\
 b_i&=\int_{-d}^d\phi_i r_G\dd u.
\end{align}
The coefficient vector $\bm U$ obeys $W\bm U''+A\bm U=\bm b e^{\ii ks}$,
with primes denoting $s$ derivatives. The generalized eigenproblem
$A\mathsf V=W\mathsf V\mathsf B^2$, $\mathsf V^T W\mathsf V=I$,
gives the circular-guide modes as columns of $\mathsf V$, with
longitudinal wave numbers $\beta_\nu$ in
$\mathsf B=\operatorname{diag}(\beta_\nu)$.
With particular coefficient vector $\bm U_p=(A-k^2W)^{-1}\bm b$, the
solution is
\begin{equation}
 \bm U(s)=\bm U_p e^{\ii ks}
 +\mathsf V[e^{\ii\mathsf B s}\bm c_+
             +e^{\ii\mathsf B(L_b-s)}\bm c_-].
\end{equation}
Here $\bm c_+$ and $\bm c_-$ are the forward and backward amplitudes
in the bend eigenbasis, launched from its entrance and exit, respectively.
Both directions are retained, with decaying
exponentials for evanescent modes. Define
$D_0=\operatorname{diag}(\beta_{nm})$ for the straight guide and
$d_i=k\int_{-d}^d\phi_i(1-h^{-1})G\dd u$.
Continuity of the total field and its physical normal derivative,
$h^{-1}\partial_s$ on the circular side, gives
\begin{align}
 W\bm U'(0)+\ii D_0\bm U(0)&=\ii\bm d,\nonumber\\
 W\bm U'(L_b)-\ii D_0\bm U(L_b)&=\ii\bm d e^{\ii kL_b}.
\end{align}
Solving these equations gives the bend amplitudes $\bm c_\pm$ and
hence $\bm U(L_b)$. The latter supplies the outgoing straight-guide
coefficients in Eq.~\eqref{eq:outgoingamplitudes}; the propagating
coefficients of both families form $\bm a_{\rm Helm}$.
Setting $U=e^{\ii ks}v$ and neglecting $v_{ss}$ gives the paraxial
system $2\ii kW\bm v'+(A-k^2W)\bm v=\bm b$.
Its entrance condition $\bm v(0)=0$ replaces the bidirectional matching.
Powers are evaluated with Eq.~\eqref{eq:portflux}.

\subsection{Branching geometry}
A source of finite transverse width must preserve current conservation
as it follows the orbit. Let $Z_b(X)$ describe the orbit and
$s'=\dd s/\dd X=\sqrt{1+(Z_b')^2}$ its arc-length derivative.
Primes in this source definition denote $X$ derivatives. For the
normalized Gaussian
$g_\sigma(x)=\exp[-x^2/(2\sigma^2)]/(\sqrt{2\pi}\sigma)$, where
$\sigma$ is its rms width, we use
\begin{align}
 J_{X,m}&=I_\omega w_m g_{\sigma_Z}(Z-Z_b)e^{\ii k s(X)},\nonumber\\
 J_{Z,m}&=Z_b'J_{X,m},\qquad \rho_m=(s'/c)J_{X,m},\nonumber\\
 w_m&=\frac{2}{b}(-1)^{(m-1)/2}e^{-\alpha_m^2\sigma_Y^2/2}.
 \label{eq:maxwellsource}
\end{align}
Here $w_m$ is the vertical Gaussian coefficient, $\rho_m$ is charge
density, and $J_{X,m}$ and $J_{Z,m}$ are current-density components.
These expressions satisfy $\partial_XJ_X+\partial_ZJ_Z=\ii\omega\rho$
exactly. We use $\sigma_Y=80\,\mu$m; $\sigma_Z$ is specified beside each
result and is measured in global-$Z$ slices.

The beam field subtracted at each port must correspond to the same
finite-width source. To express this source on a transverse port,
consider a straight segment at angle $\theta$ to the global $X$ axis,
with port center
$(X_p,Z_p)$ at orbit length $s_p$. On the plane normal to its axis,
$X=X_p+u\sin\theta$ and $Z=Z_p-u\cos\theta$. Thus
$Z-Z_b(X)=-u/\cos\theta$ and $s(X)=s_p+u\tan\theta$.
Define the rms width normal to the axis as $\sigma_u=\sigma_Z\cos\theta$.
After factoring out the common phase $e^{\ii ks_p}$, the transverse
source contains
\begin{equation}
 g_{\sigma_u}(u)\exp(\ii ku\tan\theta).
 \label{eq:sourcephase}
\end{equation}
In the charge density, the Gaussian normalization factor $\cos\theta$
cancels $s'=1/\cos\theta$. We retain the complex factor in
Eq.~\eqref{eq:sourcephase} in the particular field at the tilted main port. The
E projection uses the conserved charge density; the H
projection uses the current curl, including the transverse phase
derivative. Thus the port beam field is computed from the same finite-width
source as the interior field.

At a port of normal width $W$, the orthonormal modes $\psi_n$ are sines
for E ($n\geq1$) and cosines for H ($n\geq0$). The outgoing
Dirichlet-to-Neumann (DtN) condition is
$\partial_nF^{\rm rad}=\ii T_mF^{\rm rad}$, where
$F^{\rm rad}=F-F^{\rm beam}$ is the radiation field after subtraction
of the particular beam field $F^{\rm beam}$. The map satisfies
$T_m\psi_n=\beta_{nm}\psi_n$,
where $\beta_{nm}=\sqrt{k^2-\alpha_m^2-(n\pi/W)^2}$ is real positive
for propagation and has positive imaginary part for evanescent decay.

Let $\Omega$ be the horizontal vacuum domain, $\Gamma$ a port boundary,
and $\phi_i$ the FE basis functions. The horizontal gradient is
$\nabla=(\partial_X,\partial_Z)$. We assemble
$K_{ij}=\int_\Omega\nabla\phi_i\cdot\nabla\phi_j\dd A$,
$M_{ij}=\int_\Omega\phi_i\phi_j\dd A$, and
$(P_\Gamma)_{in}=\int_\Gamma\phi_i\psi_n\dd\ell$. For the coefficient
vector $\bm F$ of nodal values of $F$, the scalar problem is
\begin{align}
 A&=K-(k^2-\alpha_m^2)M\nonumber\\
 &\quad-\ii\sum_\Gamma
  P_\Gamma\operatorname{diag}(\beta_{nm})P_\Gamma^T,\nonumber\\
 A\bm F&=\bm f+\sum_\Gamma\bm g_\Gamma.
\end{align}
Here $A$ is the assembled system matrix, $\bm f$ is the volume source
load, and $\bm g_\Gamma$ is a load on port $\Gamma$. The volume loads are
\begin{align}
 f_i^E&=-\frac{\alpha_m}{\epsilon_0}\int_\Omega\phi_i\rho_m\dd A,\nonumber\\
 f_i^H&=\int_\Omega(\partial_X\phi_iJ_{Z,m}-\partial_Z\phi_iJ_{X,m})\dd A.
\end{align}
In the H load, integration by parts avoids differentiating the narrow
Gaussian source numerically.
At a beam port, the boundary load is
$(\bm g_\Gamma)_i=\int_\Gamma\phi_i g_\Gamma\dd\ell$, where
$g_\Gamma=\partial_nF^{\rm beam}-\ii T_mF^{\rm beam}$,
with normal derivative $-\ii kF^{\rm beam}$ at the inlet and
$+\ii kF^{\rm beam}$ at the main exit.
The particular field is computed from the same finite-width source:
for $q_n=n\pi/W$, its envelope coefficient is
$-\int_\Gamma\psi_n S_m\dd\ell/(q_n^2+\alpha_m^2)$, with $S_m$ the
scalar Helmholtz source after removal of the local carrier.
Separate quadrature panels resolve the beam core and outer port.
We check the source projection with the analytic Gaussian integrals below.
At the beam-free outlet, $g_\Gamma=0$ unless radiation is incident.

The port integrals can be checked analytically. With phase wave number
$\gamma=k\tan\theta$ and any Fourier wave number $q$,
\begin{equation}
 \int_{-\infty}^{\infty}g_{\sigma_u}(u)e^{\ii\gamma u}e^{\pm\ii qu}\,\dd u
       =e^{-\sigma_u^2(\gamma\pm q)^2/2}.
\end{equation}
Combining the two signs and restoring the port-center phase gives
the sine and cosine projections. Integration by parts gives the H
projection; its constant horizontal mode vanishes when the negligible
Gaussian tails at the walls are omitted. These expressions check the
source projection and quadrature.

To include the known phase in the FE basis, we extend $s$ along
lines normal to the orbit. Its gradient is $\bm p=\nabla s=\bm t/h$ in the
bend and $\bm p=\bm t$ in the straight continuations, where $\bm t$ is
the unit orbit tangent. For trial envelope $\widetilde F=e^{-\ii ks}F$
and test function $\varphi$, the volume bilinear form is
\begin{align}
 a(\widetilde F,\varphi)=\int_\Omega\big[&
       \nabla\widetilde F\cdot\nabla\varphi^*\nonumber\\
 &+\ii k(\widetilde F\bm p\cdot\nabla\varphi^*
                  -\varphi^*\bm p\cdot\nabla\widetilde F)\nonumber\\
 &+\{k^2(|\bm p|^2-1)+\alpha_m^2\}
                         \widetilde F\varphi^*\big]\dd A.
\end{align}
No derivative of $\widetilde F$ has been dropped. The port trace matrix becomes
$\widetilde P_{in}=\int_\Gamma e^{\ii ks}\phi_i\psi_n\dd\ell$ and the boundary
  matrix is
$-\ii\widetilde P^*\operatorname{diag}(\beta_{nm})\widetilde P^T$.
The E load replaces $\phi_i$ by $e^{-\ii ks}\phi_i$; in the H load it replaces
  $\nabla\phi_i$ by $e^{-\ii ks}(\nabla\phi_i-\ii k\bm p\phi_i)$. The original
  physical field and its gradient are reconstructed as $F=e^{\ii ks}\widetilde
F$
  and $\nabla F=e^{\ii ks}(\nabla\widetilde F+\ii k\bm p\widetilde F)$.
  This is a change of basis, not a paraxial approximation; both
formulations solve the same Maxwell problem, although their numerical
dispersion differs on a finite mesh.

The outgoing radiation flux is given by Eq.~\eqref{eq:portflux}, with
$a$ replaced by the local port width $W$. The coefficients are taken
from the Maxwell field after subtraction of the particular beam field.
Evanescent modes contribute no real outgoing flux but remain in the
DtN boundary condition. For $\beta_{nm}\simeq k$ and $\alpha_m/k\ll1$,
the flux divided by $|I_\omega|^2$ reduces to the paraxial norm
in Eq.~\eqref{eq:paraxialpower}.
At the outlet, projection requires the field on a plane normal to that
guide's own axis, with the carrier phase restored.

For comparison with the paraxial solution, we interpolate its field
onto each port from the saved transverse sections, restore
$e^{\ii ks}$, and project onto the port modes. We split quadrature at
element and section crossings to resolve the piecewise-polynomial
field, and test both the quadrature and the spacing of saved sections.
The outgoing-mode weights give a common power measure, but do not
make the paraxial field satisfy the exact relation between a Maxwell
field and its normal derivative.

To check the balance between work done by the beam and outgoing
power, we reconstruct the horizontal electric-field coefficients,
\begin{align}
 E_{X,m}&=\frac{-\ii\omega\mu_0\partial_ZF_m^H
-\alpha_m\partial_XF_m^E-\ii\omega\mu_0J_{X,m}}{k^2-\alpha_m^2},\nonumber\\
 E_{Z,m}&=\frac{\ii\omega\mu_0\partial_XF_m^H
-\alpha_m\partial_ZF_m^E-\ii\omega\mu_0J_{Z,m}}{k^2-\alpha_m^2}.
\end{align}
The horizontal magnetic coefficients are
$H_{X,m}=(-\partial_ZF_m^E+\alpha_m E_{Z,m})/(\ii\omega\mu_0)$ and
$H_{Z,m}=(\partial_XF_m^E-\alpha_m E_{X,m})/(\ii\omega\mu_0)$.
The time-averaged power supplied by the beam is
\begin{equation}
 W_{\rm beam}=-\frac b4\operatorname{Re}\sum_m\int_\Omega
 (E_{X,m}J_{X,m}^*+E_{Z,m}J_{Z,m}^*)\dd A.
\end{equation}
Let $\bm n=(n_X,n_Z)$ be the outward unit port normal. Suppressing
the vertical mode index, the full outgoing port flux integrates
$b\operatorname{Re}[n_X(F^EH_Z^*-E_Z(F^H)^*)
+n_Z(E_X(F^H)^*-F^EH_X^*)]/4$, summed over modes.
  This balance uses the total field, including the particular beam
field, unlike the radiation powers calculated after beam-field subtraction.

We eliminate the three interior unknowns of each P4 triangle before
the global solve and recover them afterwards. Mesh refinement, source
quadrature, port-mode truncation, and power balance are checked separately. The beam-driven outlet comparison retains 384 horizontal
port modes per family. Its source-region spacing is reduced from
150 to 75 $\mu$m at fixed 1.2 mm bulk spacing, and $\sigma_Z$ from
50 to 25 $\mu$m. The largest changes in branch powers are about
0.3\% and 0.7\%, respectively. These refinements assess numerical
and source-width sensitivity; they do not extrapolate to a line source.

For the local scattering test, we retain the junction near the nose
at $s=s_*$ and attach a uniform inlet at $s_{\rm in}=s_*-10$ mm.
A separate paraxial P2 calculation supplies the field from the full bend
and opening up to this plane. Subtracting the straight-guide beam
field for the width at that plane defines $v_{\rm in}^j(u)$.
This is the prescribed input to the shortened problem, not a separately
conserved part of the radiation generated in the bend.

Here $j=E,H$ identifies the scalar family and the fixed vertical
index is suppressed. Let $x_p(u)$ be the distance from one wall along
the input port. The incident coefficients are
\begin{equation}
 a_{{\rm in},n}^j=e^{\ii ks_{\rm in}}\int_{u_-}^{u_+}
              \psi_n[x_p(u)]v_{\rm in}^j(u)\dd u.
\end{equation}
We evaluate the projection element by element and retain only
propagating incident modes. The same complex coefficients, with the
carrier removed, are used at the paraxial inlet; no phases or relative
E/H amplitudes are adjusted. The Maxwell solution also gives the
reflected power.

The attached inlet creates an artificial corner with the inclined wall.
For this modified geometry, the Maxwell solution on the 0.14 mm P4 mesh
gives about 1.3\% reflected power. The branch powers differ by less
than 0.1\% between methods, but doubling the port basis to include all
propagating exit modes changes a branch power by about 0.5\%.

\newcommand{\PaechThesisURL}{%
https://tuprints.ulb.tu-darmstadt.de/946/%
1/Dissertation_Andreas_Paech_Seite_i-74.pdf}

\end{document}